\documentclass{aa}
\input{gridline_fig.sty}

\usepackage{booktabs}
\usepackage[colorlinks=true,allcolors=blue]{hyperref}
\usepackage{graphicx}
\usepackage{float}
\usepackage{lscape}
\usepackage{multirow}

\usepackage{txfonts}

\begin{document}

   \title{The Close AGN Reference Survey (CARS)}
   \subtitle{Long-term spectral variability study of the changing look AGN Mrk 1018}
    \titlerunning{Mrk 1018}
    \authorrunning{T. Saha et al.}
    \author{T. Saha \inst{\ref{aip},\ref{camk}}
    \and M. Krumpe \inst{\ref{aip}}
    \and A. Markowitz \inst{\ref{camk}}
    \and M. Powell \inst{\ref{aip}}
    \and G. Leung \inst{\ref{ut},\ref{mit}}
    \and F. Combes \inst{\ref{cnrs}}
    \and R. E. McElroy \inst{\ref{uqueensland}}
    \and J. S. Elford \inst{\ref{cardiff}}
    \and M. Gaspari \inst{\ref{modena}}
    \and N. Winkel \inst{\ref{mpia}}
    \and A. L. Coil \inst{\ref{ucsd}}
    \and T. Urrutia \inst{\ref{aip}}
    }

      \institute{Leibniz-Institut für Astrophysik Potsdam, An der Sternwarte 16, 14482 Potsdam, Germany \label{aip}
      \and Nicolaus Copernicus Astronomical Center of the Polish Academy of Sciences, ul.\ Bartycka 18, 00-716 Warszawa, Poland\label{camk}
      \and Department of Astronomy, The University of Texas at Austin, 2515 Speedway, Stop C1400, Austin, Texas 78712-1205, US\label{ut}
      \and MIT Kavli Institute for Astrophysics and Space Research, 77 Massachusetts Ave., Cambridge, MA 02139, USA \label{mit}
      \and Observatoire de Paris, LUX, Collège de France, CNRS, PSL University, Sorbonne University, 75014, Paris, France\label{cnrs}
      \and Centre for Astrophysics, University of Southern Queensland, 37 Sinnathamby Boulevard, Springfield, Qld 4300, Australia \label{uqueensland}
      \and Instituto de Estudios Astrof\'{\i}sicos, Facultad de Ingenier\'{\i}a y Ciencias, Universidad Diego Portales, Av. Ej\'ercito Libertador 441, Santiago, Chile \label{cardiff}
      \and Department of Physics, Informatics and Mathematics, University of Modena and Reggio Emilia, 41125 Modena, Italy \label{modena}
      \and Max-Planck-Institut f{\"u}r Astronomie, K{\"o}nigstuhl 17, D-69117 Heidelberg, Germany \label{mpia}
      \and Department of Astronomy and Astrophysics, University of California San Diego, La Jolla, CA 92093, USA \label{ucsd}
      }

   \date{Received ; accepted }
   
   \abstract{Changing-look AGNs (CLAGN) are accreting supermassive black hole systems that undergo variations in optical spectral type, driven by major changes in accretion rate.  Mrk 1018 has undergone two transitions, a brightening event in the 1980s and a transition back to a faint state over the course of 2–3 years in the early 2010s.}
   {We characterize the evolving physical properties of the source's inner accretion flow, particularly during the bright-to-faint transition, as well as the morphological properties of its parsec-scale circumnuclear gas.}
   {We model archival X-ray spectra from {\it XMM-Newton}, {\it Chandra}, {\it Suzaku}, and {\it Swift}, using physically-motivated models to characterize X-ray spectral variations and track Fe K$\alpha$ line flux.  We also quantify Mrk 1018's long-term multi-wavelength spectral variability from optical/UV to the X-rays.}
   {Over the duration of the bright-to-faint transition,
   the UV and hard X-ray flux fell by differing factors, roughly 24 and 8, respectively. The soft X-ray excess faded, and was not detected by 2021.  In the faint state, when the Eddington ratio drops to $\log (L_{\rm bol}/L_{\rm Edd})\lesssim -1.7$, the hot X-ray corona photon index shows a ‘softer-when-fainter’ trend, similar to that seen in some black hole X-ray binaries and samples of low-luminosity AGNs. Finally, the Fe K$\alpha$ line flux has dropped by only half the factor of the drop in the X-ray continuum.}
   {The transition from the bright state to the faint state is consistent with the inner accretion flow transitioning from a geometrically-thin disk to an ADAF-dominated state, with the warm corona disintegrating or becoming energetically negligible, while the X-ray-emitting hot flow becoming energetically dominant.  Meanwhile, narrow Fe K$\alpha$ emission has not yet fully responded to the drop in its driving continuum, likely because its emitter extends up to roughly 10 pc.}
      

   \maketitle

\section{Introduction}\label{sec:intro}
Active galactic nuclei (AGN) are powered by accretion onto supermassive black holes \citep{soltan1982} and can emit from the radio band up to gamma rays.  Based on its optical spectrum, a Seyfert AGN can be classified into two general types, namely, type 1, where broad emission lines (FWHM $\gtrsim$ 1000~km~s$^{-1}$) and narrow forbidden emission lines are present, and type 2 where broad emission lines are absent and only narrow forbidden emission lines can be observed. Intermediate types \citep[1.2, 1.5, 1.8, and 1.9;][]{osterbrock1976} span an intermediate range of broad line properties, e.g. type 1.9 sources exhibit a broad H$\alpha$ emission line, but broad H$\beta$ is not detected.

Accretion onto compact objects is generally observed to be a dynamic, hence temporally variable process. Persistently accreting Seyfert AGN exhibit stochastic variability up to a factor of $\sim$10--20 in the X-rays \citep[e.g.][]{markowitz2004} and from a few percent and up to a factor of $\sim$five in the optical \citep[e.g.][]{uttley2004} on timescales ranging from a few days to few years. A leading mechanism behind stochastic variability is inwardly-propagating fluctuations in local mass accretion rate \citep{ingram2011}, induced by e.g., fluctuations associated with the magneto-rotational instability \citep{balbus1991}. The community has accumulated over 200 observations of Seyferts and quasars that exhibit extreme continuum variability, significantly in excess of the ‘normal’ stochastic variability, accompanied by changes to their optical spectral type. Specifically, the optical spectral type changes as one or more broad emission lines appear or disappear. Recent select examples include 1ES~1927+654 \citep{ricci2020}, Mrk~590 \citep{denney2014}, and NGC~2617 \citep{shappee2014}. In most cases, optical spectral changes are driven by changes in the underlying ionizing continuum luminosity rather than in line-of-sight obscuration \citep{korista1995, korista1997, korista2000, wu2023}. Results generally support that the prime mechanism driving these continuum luminosity variations is large changes in accretion rate \citep[e.g.][]{lamassa2015, zeltyn2024, ricci2023}, due to e.g., major changes in the structure of the accretion flow \citep[e.g.,][]{noda2018}, instabilities originating in the disk \citep[e.g.][]{lightman1974}, or tidal disruption events (TDEs) occurring in AGN  \citep{trakhtenbrot2019,homan2023}\footnote{ The family of extreme variability AGNs also includes events associated with variations in line-of-sight X-ray obscuration, e.g., \citet{markowitz2014}, \citet{serafinelli2021}, \citet{marchesi2022}, \citet{markowitz2024}, \citet{sengupta2024} and references therein.}.

For changing look AGNs (CLAGN) driven by drastic changes in accretion rate, a broad range of behavior is observed in the X-ray and broadband continuum spectrum. Some sources exhibit drastic changes in X-ray spectral shape with rapid spectral hardening or softening alongside a decrease or increase in multi-band flux. For example, \citet[][]{ricci2020} report extreme X-ray spectral softening in the CLAGN 1ES 1927+654 due to the drastic weakening of X-rays above $\sim$2.0~keV, attributed to the destruction of the hard X-ray emitting hot corona. 
After the CLAGN transition in 2017, the hard X-ray-emitting
hot corona recovered \citep{ricci2020}. Subsequent multi-wavelength
monitoring indicated that the accretion flow
transitioned from a slim disk to a thin disk
two years after the outburst \citep{li2024}.
Meanwhile, \cite{ghosh2022} report rapid variability in hot-corona X-ray photon index $\Gamma$ during the long-term flaring in the CLAGN source Mrk 590. In contrast, some other CLAGN exhibit large changes in flux with no significant detectable changes in spectral shape \citep[e.g.][]{lamassa2015, kollatschny2023, saha2023}. Furthermore, over a dozen sources (so far) are also known to exhibit repeat CLAGN events \citep{wang2024}. 
All these sources demonstrate that changing look AGN transitions induce a broad range of divergent changes in the behavior of the multiwavelength spectra.
Characterizing the spectral behavior of each individual source helps reveal the unique characteristics of their inner accretion flow.

Markarian 1018 (Mrk 1018 hereafter) is a galaxy merger remnant system that hosts an AGN at its center. The AGN was initially classified to be of type 1.9 \citep{osterbrock1981}, which subsequently changed its type twice: (a) type 1.9 to type 1 by 1984 \citep{cohen1986} and (b) type 1 to type 1.9 detected in 2016 \citep{mcelroy2016}. The later transition was accompanied by a drop in optical flux. X-ray, optical, and UV monitoring of the source over the last two decades have revealed that all wavelengths experienced a quasi-simultaneous flux drop (between 2010--2013 as shown later in this work) associated with the changing look transition of 2016 \citep{lyu2021, husemann2016}. An absence of extrinsic line-of-sight change in the faint state X-ray spectrum \citep{husemann2016, lamassa2017}, along with a lack of change in fractional polarization \citep{hutsemekers2020}, have ruled out a changing obscuration as the origin of the transition. Consequently, on the basis of broadband optical-UV-X-ray studies, \cite{lyu2021} and \citet{noda2018} have demonstrated that the extreme multiwavelength variability was triggered by a drop in accretion rate and the accretion flow has undergone a structural change-- transition from a geometrically-thin optically-thick accretion disk to a radiatively-inefficient flow. \cite{noda2018} explore the possibility of a radiation pressure instability \citep[e.g.][]{szuszkiewicz2001} or hydrogen ionization instability. Mrk 1018, being a post-merger system, is thought to be associated with abundant cold gas. This environmental factor strengthens the plausibility of some external fueling bursts, such as chaotic cold accretion (CCA) precipitation \citep{gaspari2013}. A recent archival study by \citet{veronese2024} also explored alternative pathways such as dynamical friction, external multi-phase inflows and CCA \citep{gaspari2013,gaspari2017,gaspari2020} as mechanisms for triggering extreme variability in Mrk 1018.

Mrk 1018 was observed to have exhibited two subsequent outbursts after the major flux-drop event post 2013: once between October 2016 and February 2017 \citep[$\Delta U\simeq0.4$,][]{krumpe2017} with optical broad emission line change in H$\alpha$ and a second time in mid 2020 \citep[$\Delta u'=2.71$,][]{brogan2023} with significant optical broad (H$\beta$ and H$\alpha$) emission line flux and profile variation within several months \citep{lu2025}. Further X-ray spectroscopic studies have been undertaken to compare the Fe~K$\alpha$ line properties in the pre- and post- shutdown phases. A significant change in the Fe K$\alpha$ equivalent width is also reported in the pre- and post- shutdown phase, primarily due to the suppression of the underlying continuum \citep{lamassa2017}.

In this paper, we perform a multi-wavelength spectral decomposition focussed mostly on X-ray and UV using physically motivated models and track the time-evolution of individual spectral components of Mrk 1018 from its bright state to its faint state, as well as track its broadband spectral variability. We relate the time-dependent behavior of the multi-band emission to structural changes in the inner accretion flow. Additionally, we examine in detail the flux and profile behavior of the Fe K$\alpha$ emission line and gain insights into the circumnuclear line-emitting gas.

The remainder of the paper is structured as follows: In Sect.~\ref{sec:data-reduction}, we describe the data reduction procedure. In Sect.~\ref{sec:data-analysis}, we perform analysis of high signal/noise X-ray and broadband SED to establish the best physical model of the X-ray continuum and characterize its variability characteristics. In Sect.~\ref{sec:discussion}, we summarize the distinct changes in the spectral components inferred from our spectral analysis, and connect the physical process that triggered the CL transition, its impact on the physical properties of the inner accretion flow, and distant substructures. Finally, we summarize our conclusions in Sect.~\ref{sec:conclusions}.

Throughout, the paper all mentioned uncertainties correspond to the 90\% confidence limit. For X-ray and broadband SED fitting using Bayesian methods, the 90\% confidence range is bounded by the 0.5th and the 0.95th quantile of the given single posterior distribution, unless stated otherwise. The defined upper limits are the 0.95th quantile of the single posterior distribution.
For the purposes of this paper, the soft-X-ray band is defined as 0.3--2~keV and the hard X-ray band is defined as 2--10~keV band, unless stated otherwise.
In this paper, we assume a flat cosmology, $H_0 = 70$~km~s$^{-1}$~Mpc$^{-1}$, $\Omega_{\rm M}$=0.3, and $\Omega_{\rm vac}=0.7$. Throughout the paper, Mrk 1018's redshift is taken to be $z=0.043$ \citep{mcelroy2016}. The luminosity distance is $d_{\rm L}= 188.5$~Mpc and the angular distance is $d_{\rm A}=173.4$~Mpc.

\section{Data reduction} \label{sec:data-reduction}

\begin{table*}[h]
\caption{X-ray observations of Mrk 1018 with major X-ray facilities.}\label{tab:observation-list}
\centering
    \begin{tabular}{ccccccccccc}
    \hline
    \hline
    Date & Instrument      & Abbr. & Obs.ID & Obs. & GTI& Counts & Comments & Opt./UV obs.  \\
    yyyy-mm-dd &&&& mode & (ks) &          & & Filter bands\\
    \hline
   \multicolumn{8}{c}{\textit{bright type 1 phase}} \\\hline
   2005-01-15 & {\it XMM-Newton} & XMM1 & 0201090201 & FF,T   & 1.4 & 7177 & high variable bkg. & U,W1,M2,W2 \\
   2005-08-05 & {\it Swift} & Sw1 & 00035166001 & PC & 5.2 & 1778 & Piled-up & V,B,U,W1,M2,W2 \\
   2008-06-11 & {\it Swift} & Sw2 & 00035776001 & PC & 4.8 & 1147 & Piled-up & V,B,U,W1,M2,W2 \\
   2008-08-07 & {\it XMM-Newton} & XMM2 & 0554920301 & SW,M   & 9.2 & 61263 &  -- & U,W1,M2 \\
   2009-07-03 & {\it Suzaku}     & SUZ   & 704044010  &  --    & 43.9 & 86536 & -- & --\\
   2010-11-27 & {\it Chandra}    & C1    & 12868      & ACIS-S & 22.7 & 741 & Piled-up & --\\
   \hline
   \multicolumn{8}{c}{\textit{transition phase}} \\\hline
   2013-03-01 & {\it Swift} & Sw3 & 00049654001 & PC & 0.7 & 57  & low cts. & V,B,U,W1,M2,W2 \\
   2013-06-07 & {\it Swift} & Sw4 & 00049654002 & PC & 1.3 & 291 & low cts. & V,B,U,W1,M2,W2\\
   2014-06-09 & {\it Swift} & Sw5 & 00049654004 & PC & 2.1 & 86  & low cts. & V,B,U,W1,M2,W2\\
   \hline
   \multicolumn{8}{c}{\textit{faint type 1.9 phase}} \\\hline
   2016-02-25 & {\it Chandra}    & C2    & 18789      & ACIS-S & 27.2 & 4171 &-- & --\\
   2017-02-17 & {\it Chandra}    & C3    & 19560      & ACIS-S & 48.6 & 12778 &--& -- \\
   2018-06-12 & {\it Chandra}    & C4    & 20366      & ACIS-S & 18.2& 3252 & --& -- \\
   2018-07-23 & {\it XMM-Newton} & XMM3 & 0821240201 & FF,M   & 52.2& 38139 & -- & M2\\
   2018-09-09 & {\it Chandra}    & C5    & 20369      & ACIS-S & 19.1 & 5103 & -- & --\\
   2019-01-04 & {\it XMM-Newton} & XMM4 & 0821240301 & FF,M   & 43.4& 15925 & -- & M2 \\
   2019-02-06 & {\it Chandra}    & C6    & 21432      & ACIS-S & 22.7 & 3883 & -- & --\\
   2019-02-07 & {\it Chandra}    & C7    & 22082      & ACIS-S & 18.2& 3360 & --  & --\\
   2019-10-09 & {\it Chandra}    & C8    & 21433      & ACIS-S & 40.9 & 3773 & -- & --\\
   2021-02-04 & {\it XMM-Newton} & XMM5 & 0864350101 & FF,M   & 41.0& 14995 & after outburst & M2\\
   \hline
   \hline
   \end{tabular} 
\tablefoot{
Observing Mode: 
FF -- full frame mode (\textit{XMM-Newton} EPIC-pn), 
SW -- small window mode (\textit{XMM-Newton} EPIC-pn); 
T -- thin filter, 
M -- medium filter; 
GTI denotes the good time interval for \textit{XMM-Newton} pn pattern 0, or for \textit{Suzaku} (all XIS detectors).
Specified counts are from: \textit{XMM-Newton} 0.3--10.0~keV range (pn pattern 0 only), \textit{Chandra}: 0.5--10.0~keV range, \textit{Suzaku}: XIS0+XIS1+XIS3 in 0.4--10.0 keV range. The {\it XMM-Newton} and {\it Suzaku} observations yielded the highest count spectra when adding up all instruments (e.g., EPIC-pn pattern 0 + pn pattern 1-4 + MOS1 + MOS2). Only these observations are used for our detailed spectral fitting in Sect.~\ref{sec:xmm-suzaku-analysis}.}
\end{table*}

\subsection{{\it XMM-Newton}} \label{sec:xmm}

\subsubsection{EPIC} \label{sec:xmm-epic}
Mrk 1018 has been observed five times with {\it XMM-Newton} \citep{jansen2001} between January 2005 and February 2021: two observations (ObsID 0201090201 and 0554920301; PIs Barcons and Corral; durations of 11.9 ks and 17.6 ks, respectively) in the bright state and three observations (ObsID 082124020, 0821240301, and 0864350101; PI: Krumpe; durations of 74.8 ks, 67.7 ks, and 65.0, respectively) in the faint state. For ObsID 0554920301, EPIC-pn was operated in small window mode. All other observations and cameras used the full frame mode.

The spectra are extracted with the {\tt SAS} package (version 21.0.0), Science Analysis Software \citep{gabriel2004}, and {\tt HEASOFT} (v6.24), including the corresponding calibration files. The tasks {\tt emproc} and {\tt epproc} are used for generating linearized photon event list from the raw EPIC data. For the EPIC-pn detectors, we use circular source-free background regions, located a few arcminutes from the source region on the same chip. For the MOS detectors, we use both circular and annulus-shaped source free background regions around the source on the same CCD. Furthermore, we follow the recommended flag selection of the macros {\tt XMMEA\_EP} and {\tt XMMEA\_EM}. The low energy cut-off is set to 0.3 keV, and the high energy cut off is fixed to 12 keV. Using the task {\tt epatplot}, we verifed that the effect of photon pile-up is negligible in all {\it XMM-Newton} observations. 

We select X-ray events corresponding to pattern 0--12 (single, doubles, and triples) for the MOS detectors. For the EPIC-pn detector, we separate the spectra into single events (pattern 0) and double events (1--4). This is because in small window mode pattern 1--4 spectra should be ignored at low energies ($\lesssim$0.5 keV) and pattern 0 spectra have an improved energy resolution over pattern 1--4 events.

ObsID 0201090201 is found to exhibit a high and variable 10--13~keV background, with count rates up to $\sim$14.5~ct~s$^{-1}$. We screened against background flaring by removing those times when the EPIC-pn $E>10.0$~keV background rate exceeded 12~cts~s$^{-1}$. This cut yielded exposure times ranging between 1200 and 1500 ks in pn, MOS1, and MOS2, with a summed total of 14220 0.3--10 keV counts in pn0 + pn14 + MOS1+MOS2. We also tested more conservative background rate cuts (e.g., 3~cts~s$^{-1}$), but we found no significant impact on best-fit model parameters for the source spectra, just larger parameter uncertainties.

For ObsIDs 0554920301, 0821240201, 0821240201, and 0864350101, high background cleaning resulted in 10.0, 52.2, 43.4, \& 41.0 ks of pn integration time and 8$\times10^4$, $4.9\times10^4$, $2.0\times10^4$, and $2.0\times10^4$ total pn counts, respectively. The source spectra were extracted from circular region of radii ranging between 35--50$\arcsec$ and background spectra were extracted from circular regions of radii ranging between 80--120$\arcsec$. The details of all {\it XMM-Newton} observations are summarized in Table \ref{tab:observation-list}.

\subsubsection{Optical monitor} \label{sec:xmm-om}
The {\it XMM-Newton} optical monitor (OM) observations \citep{mason2001} in imaging mode were reduced using the \texttt{omichain} pipeline processing, which implements flat fielding, source detection, correction for detection dead time and aperture photometry, finally creating mosaiced images. We do not find any imaging artifacts near the source, and the source is always located at the center of the CCD. 

Additionally, for the last observation (ObsID: 0864350101) where Mrk 1018 is in its faint state \citep[e.g.][]{brogan2023}, we fit the brightness profiles of Mrk 1018 and the three stars in its field of view \citep[see Fig. 1 of][]{brogan2023} obtained from the UVM2 filter. 
We find that the FWHM of Mrk 1018's UV brightness profile, measured at approximately 3$\farcs$5, is closely comparable to that of three field stars (ranging from 3$\farcs$1 to 3$\farcs$2), indicating only a marginally broader spatial extent relative to the UVM2 point spread function (PSF). Since \textsc{omichain} photometry applies a PSF-based aperture correction, any filter-dependent variation in the host galaxy's surface brightness profile could introduce systematic uncertainties in the derived photometry.
However, our observations of Mrk 1018's brightness profile indicate that the AGN flux dominates significantly over the host galaxy in the UV band, indicating that any residual filter-dependent aperture correction error can be neglected.

The generated combined source list file was used along with the \texttt{om2pha} command to generate \textsc{Xspec}-readable spectral files for the filters (Table~\ref{tab:observation-list}) used in the observations.
The canned response files available on the ESA \textit{XMM-Newton} website\footnote{\href{https://sasdev-xmm.esac.esa.int/pub/ccf/constituents/extras/responses/OM/}{https://sasdev-xmm.esac.esa.int/pub/ccf/constituents/extras/\\responses/OM/}}
were used for analysis in \texttt{xspec}.

\begin{figure*}
    \centering
    \gridline{\fig{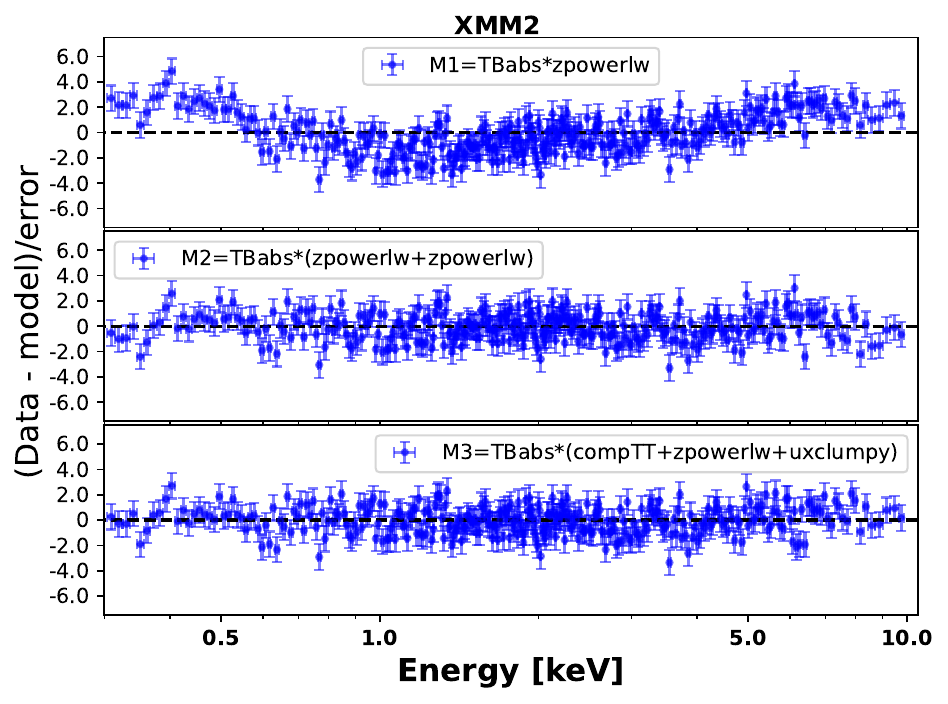}{0.48\textwidth}{(a)}
              \fig{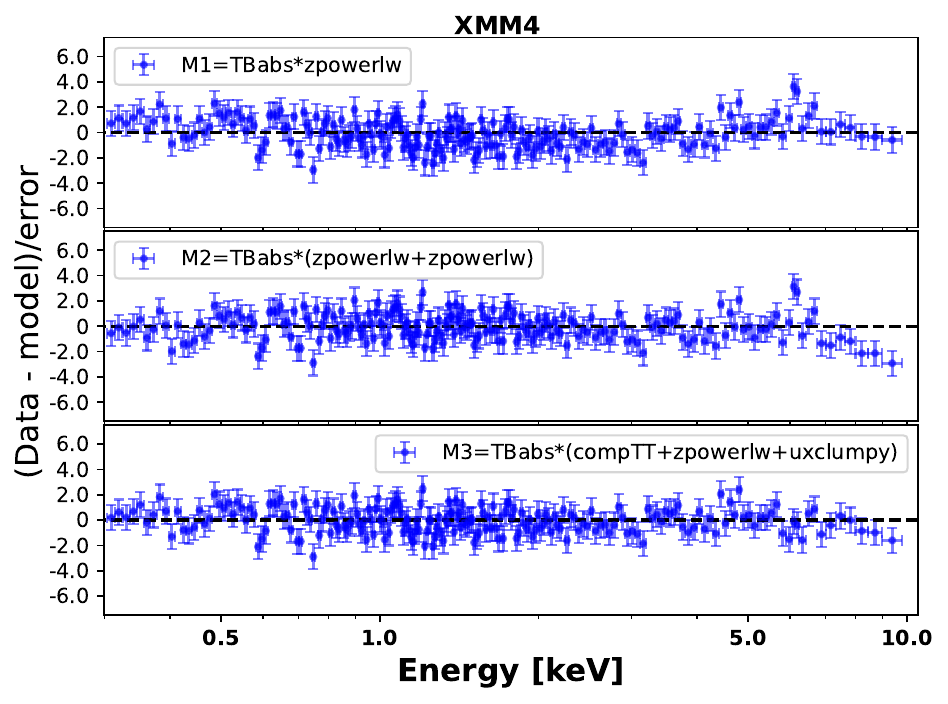}{0.48\textwidth}{(b)}
              }
    \caption{\textit{XMM-Newton} EPIC-pn pattern-0 data$-$model residuals from the X-ray spectral fitting using the models described in Sect.~\ref{sec:xmm-suzaku-analysis}. Left: bright state observation XMM2 (2008); right: faint state observation XMM4 (2019). Significant improvement in the residuals can be seen in both of the multi-component models \texttt{M2} and \texttt{M3} compared to the single power-law model, \texttt{M1}.}
    \label{fig:residuals}
\end{figure*}

\begin{figure*}
    \centering
    \gridline{\fig{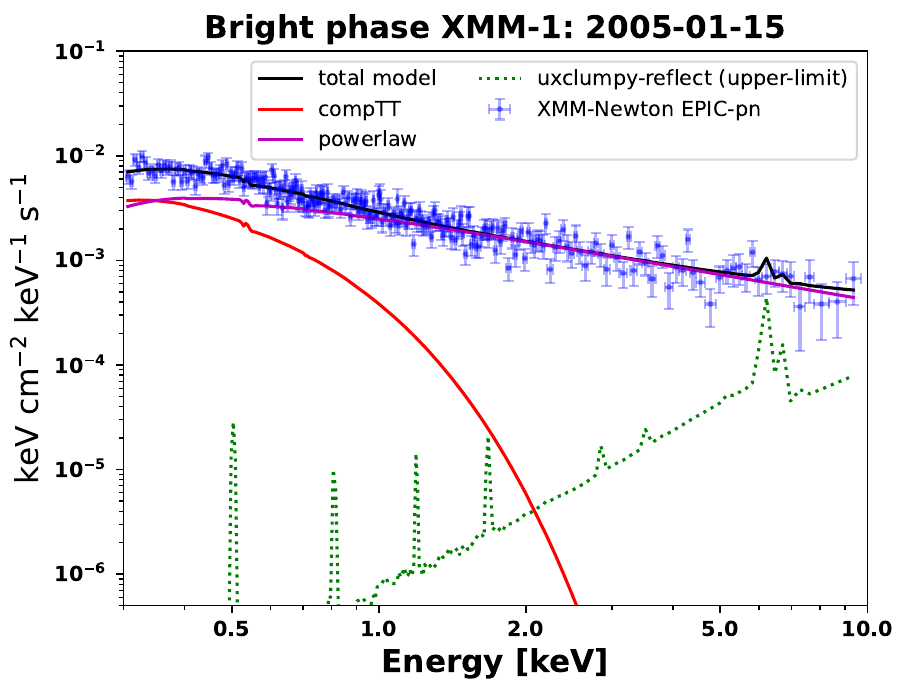}{0.3375\textwidth}{(a)}
              \fig{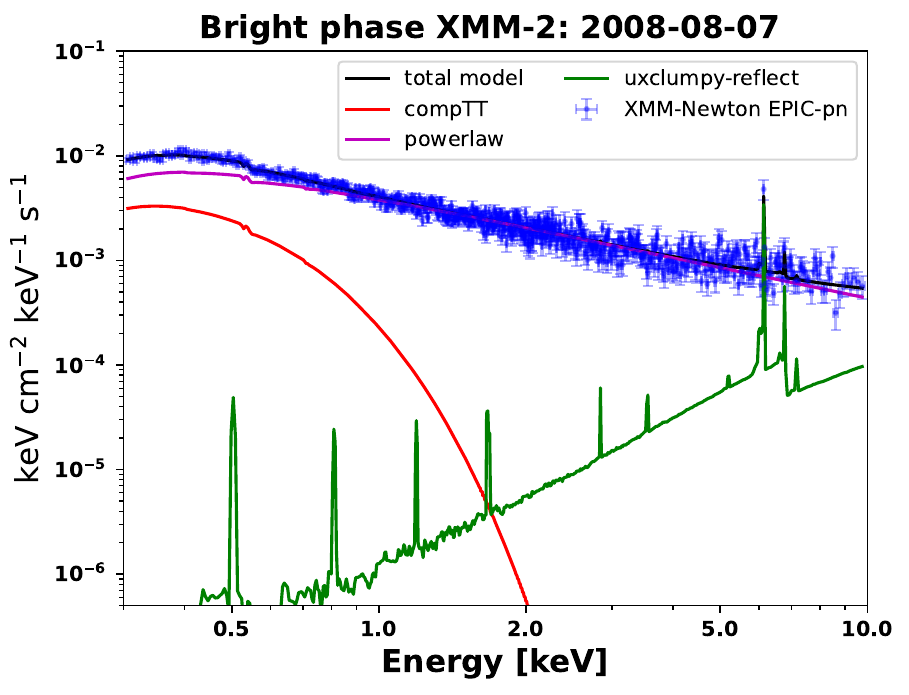}{0.3375\textwidth}{(b)}
              \fig{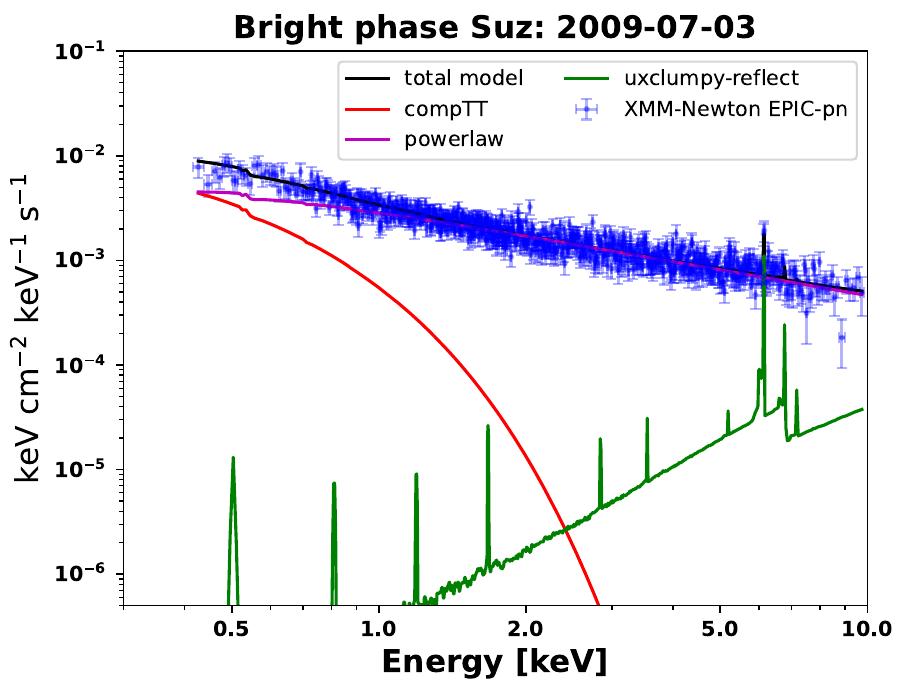}{0.3375\textwidth}{(c)}
              }

    \gridline{\fig{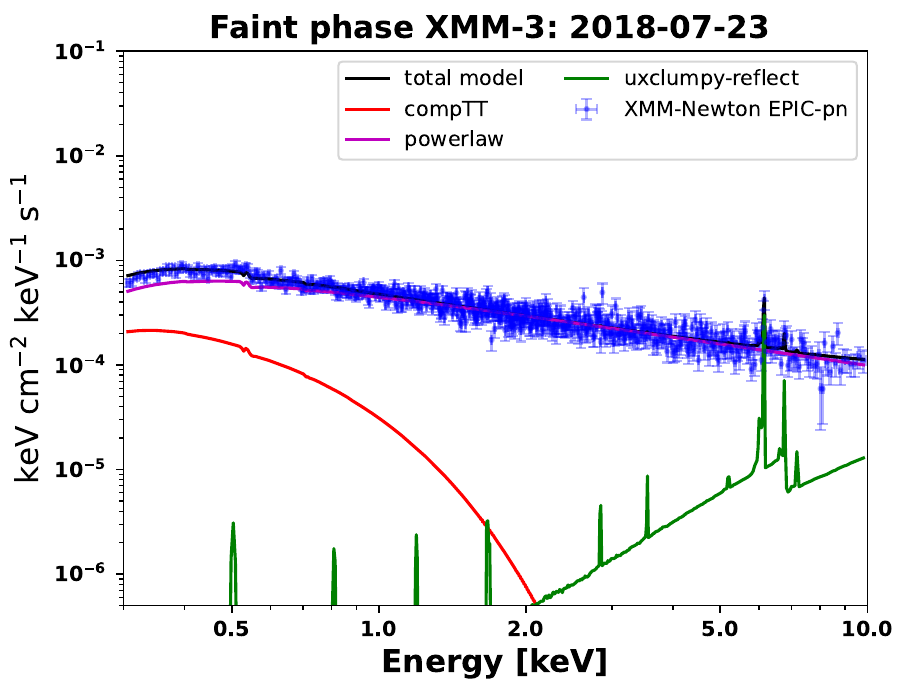}{0.3375\textwidth}{(d)}
              \fig{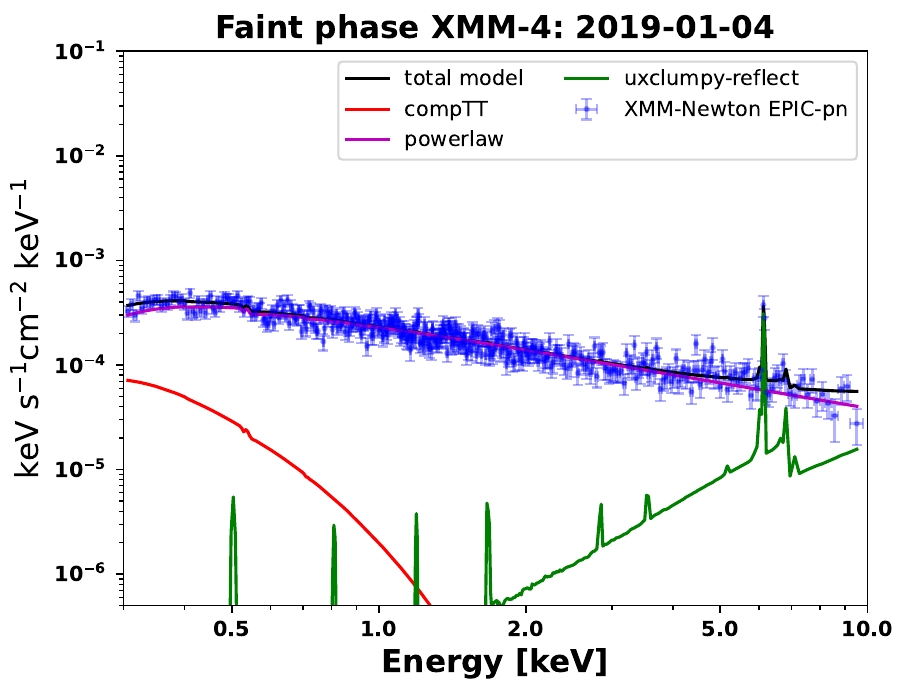}{0.3375\textwidth}{(e)}
              \fig{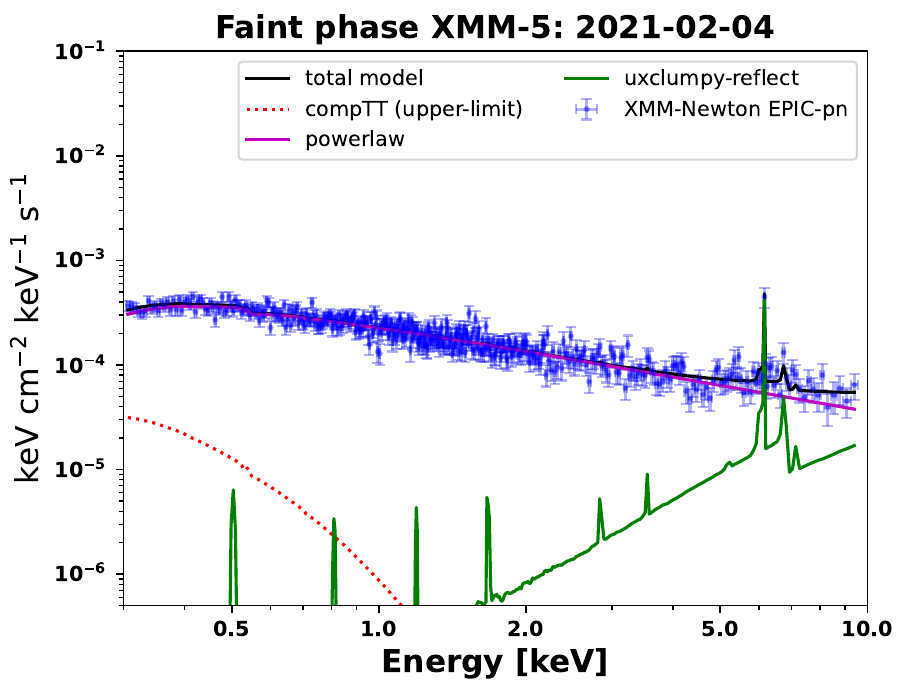}{0.3375\textwidth}{(f)}}
    
    \caption{ Unfolded best-fitting models 
    to the {\it XMM-Newton} EPIC-pn pattern 0 and {\it Suzaku} XIS-0 data. The spectral model is the best-fitting, physically motivated model \texttt{M3} (Sect.~\ref{sec:xmm-suzaku-analysis}). 
    Solid curves: model evaluated at best fit value; dotted curves: model evaluated at the upper limit; blue markers: unfolded data; red curve: soft X-ray excess modeled by \texttt{compTT}; magenta curve: hot-corona powerlaw; green curve: \texttt{uxclumpy} torus reflection model; black curve: the total unfolded X-ray spectrum.
    Panels (a), (b), and (c) show bright-state spectra; panels (d), (e), and (f) are for the faint state.
    These plots illustrate how all X-ray spectral components drop significantly in flux. The strongest drop is for the soft excess, which ultimately disappears in 2021.}
    \label{fig:xmm-suz-spectra}
\end{figure*}

\begin{table*}[hbt!]
\centering
\caption{Best parameter estimates obtained from BXA fits to the {\it XMM-Newton} and {\it Suzaku} datasets}\label{tab:xmm-suz-parameters}
\begin{tabular}{lllllll}
\hline
\hline
Parameters       & XMM1 & XMM2 & Suz & XMM3 & XMM4 & XMM5  \\
\hline
\multicolumn{7}{l}{\texttt{M1=TBabs*zpowerlw}} \\
\hline
$\Gamma_{\rm M1}$          & $2.08 \pm 0.03$ & $2.10 \pm 0.01$ & $1.94^{+0.01}_{-0.02}$ & $1.77 \pm 0.01$ & $1.77^{+0.02}_{-0.01}$ & $1.77 \pm 0.02$ \\
$A$ [$10^{-3}$]   & $3.73\pm 0.08$           & $5.13^{+0.03}_{-0.04}$   & $3.82 \pm 0.05$  & $0.57 \pm 0.01$ & $0.275 \pm 0.004$        & $0.268 \pm 0.004$ \\
$C$          & 649.32  & 2202.89  & 2952.62  & 1976.62 & 1222.69 & 1155.96  \\
$C$/dof      &  1.27   & 1.54  & 1.27  & 1.21  & 1.25    & 1.22 \\
$\rm \log Z$ & $-$148.87 & $-$487.89  & $-$649.38 & -438.98 & $-$274.40 & $-$259.81\\
\hline
\multicolumn{7}{l}{\texttt{M2=TBabs*[zpowerlw(1) + zpowerlw(2)]}}\\
\hline
$\Gamma_{\rm 1}$  & $2.68^{+0.36}_{-0.23}$ & $2.46^{+0.09}_{-0.07}$ & $3.18^{+0.23}_{-0.19}$ &$1.97^{+0.08}_{-0.05}$   & $2.03^{+0.09}_{-0.06}$ &  $1.96 \pm 0.03$          \\
$A_1$ [$10^{-3}$] & $2.02^{+0.83}_{-0.71}$ & $3.92^{+0.36}_{-0.46}$ & $1.43^{+0.32}_{-0.28}$ &$0.50^{+0.03}_{-0.05}$   & $0.24^{+0.014}_{-0.025}$ & $0.244^{+0.005}_{-0.007}$            \\
$\Gamma_{\rm 2}$  & $1.39^{+0.24}_{-0.36}$ & $1.21^{+0.17}_{-0.20}$ & $1.64^{+0.05}_{-0.06}$ &$0.82^{+0.25}_{-0.24}$  
    & $0.77^{+0.24}_{-0.23}$ & $0.58^{+0.16} _{-0.07}$             \\
$A_2$ [$10^{-3}$] & $1.39^{+0.70}_{-0.77}$ & $0.90^{+0.44}_{-0.33}$ & $2.44^{+0.27}_{-0.29}$ & $0.054^{+0.052}_{-0.025}$ & $0.028^{+0.023}_{-0.012}$  & $0.016^{+0.007}_{-0.003}$          \\
$C$         & 533.06    & 1552.48  & 2390.05 & 1748.31  & 1079.43  & 1039.99 \\
$C$/dof     & 1.04      & 1.09     &  1.03 & 1.07  & 1.10  & 1.10 \\
$\rm \log Z$ & $-$125.78  & $-$349.69 & $-$530.01 & $-$392.20 & $-$246.03 & $-$238.62\\

\hline
\multicolumn{7}{l}{\texttt{M3=TBabs*(compTT + zpowerlw + uxclumpy-reflect)}} \\
\hline
$k_{\rm B}T_{\rm e, warm}$ & $0.18 \pm 0.04$ $^{\dag}$ & $0.13^{+0.02}_{-0.03}$ & $0.20^{+0.03}_{-0.04}$ & $0.20^{+0.03}_{-0.02}$ & $0.17^{+0.09}_{-0.11}$ $^{\dag}$ & $0.17$*\\
$\tau_{\rm warm}$  & $18^{+6}_{-3}$ $^{\dag}$   & $28^{+12}_{-4}$  & $15^{+5}_{-2}$ & $18^{+2}_{-3}$ & $14^{+12}_{-5}$ $^{\dag}$ & $14$*\\
$A_{\rm compTT}$   & $1.44^{+1.09}_{-0.62}$ $^{\dag}$ & $0.75^{+0.25}_{-0.40}$ &  $2.60^{+0.98}_{-1.09}$ & $0.057^{+0.024}_{-0.011}$ & $0.12^{+0.52}_{-0.10}$ & $<0.1$ \\
$\Gamma$           & $1.80^{+0.11}_{-0.08}$ & $1.97 \pm 0.03$ & $1.82^{+0.03}_{-0.02}$ & $1.68^{+0.02}_{-0.03}$ & $1.80^{+0.03}_{-0.03}$ & $1.83^{+0.02}_{-0.03}$\\
$A_{\rm PL}$ [$10^{-3}$] & $2.85^{+0.23}_{-0.22}$ & $4.4 \pm 0.11$   & $3.28^{-0.12}_{+0.10}$ & $0.51^{+0.01}_{-0.02}$ & $0.26 \pm 0.01$ & $0.261 \pm 0.005$\\
$A_{\rm REFL}$ [$10^{-3}$] & <9.6           & $15.9^{+4.7}_{-3.7}$   & $4.40^{+1.61}_{-1.27}$ & $1.06^{+0.21}_{-0.26}$ & $1.85^{+0.40}_{-0.30}$ & $2.15^{+0.39}_{-0.36}$\\
$\left[ \frac{F_{\rm SX,COMPTT}}{F_{\rm TOT}} \right]_{\rm 0.3-2.0~keV}$ & $0.25^{+0.05}_{-0.06}$ & $0.14 \pm 0.02$ & $0.24 \pm 0.03$ & $0.10^{+0.02}_{-0.01}$ & $0.03 \pm 0.02$ & $<0.02$\\
$\left[ \frac{F_{\rm REFL,UXCL}}{F_{\rm TOT}} \right]_{\rm 2-10~keV}$ & <0.07 & $0.08\pm0.02$ & $0.04 \pm 0.01$ & $0.06 \pm 0.01$ & $0.16 \pm 0.02$ & $0.18 \pm 0.03$\\
$C$     & 531.78 & 1555.71 & 2330.62 & 1701.94  & 1031.37 & 940.33 \\
$C$/dof  & 1.04 & 1.09 & 1.00 & 1.05 & 1.06 & 0.99 \\
$\log Z$ & $-$127.16 & $-$353.62 & $-$520.17 & $-$388.63 & $-$235.51 & $-$216.20\\
\hline
\hline 

\end{tabular}
\tablefoot{ All parameter estimates are the median value of the BXA-posterior distributions, and the uncertainties correspond to 90\% confidence limits, unless specially marked. \\
$\dag$Exhibits bimodal posteriors. The quoted parameter values correspond to the distribution around the mode about which the integrated probability is higher.\\
$*$Parameters were kept frozen when calculating the upper limit to $A_{\rm compTT}$.\\
}
\end{table*}

\subsection{{\it Suzaku}} \label{sec:suzaku-xis}
\textit{Suzaku} observed Mrk~1018 on 3 July 2009 (ObsID 704044010). Data were obtained with the X-ray imaging CCDs (XIS; \citealt{koyama2007}), which consist of three front-illuminated (FI) CCDs and one back-illuminated (BI) CCD, although one of the FI CCDs (XIS2) was not used. Data were also obtained using the non-imaging hard X-ray detector (HXD; \citealt{takahashi2007}); the target was placed at the ``HXD nominal'' position. The XIS and HXD data are processed with {\em Suzaku} pipeline processing version 2.0.6.13. The XIS data reduction follow the {\em Suzaku} Data Reduction Guide\footnote{\href{http://heasarc.gsfc.nasa.gov/docs/suzaku/analysis/abc/abc.html}{http://heasarc.gsfc.nasa.gov/docs/suzaku/analysis/abc/abc.html}}. As per a notice for a calibration database update\footnote{\href{http://heasarc.gsfc.nasa.gov/docs/suzaku/analysis/sci\_gain\_update.html}{http://heasarc.gsfc.nasa.gov/docs/suzaku/analysis/sci\_gain\_update.html}} the task {\tt xispi} is run on the unfiltered events again. The events are filtered using standard extraction criteria (e.g., avoiding the South Atlantic Anomaly, low Earth elevation angles, etc.). The XIS good integration time is 44 ks per CCD. We extract the source events using a region 2$\arcmin$ in radius, the background events using four regions each 1$\arcmin$ in radius located several arcminutes from the source, also avoiding the $^{55}$Fe calibration source events. The response (RMF) and auxiliary (ARF) file are produced using the tasks {\tt xisrmfgen} and {\tt xissimarfgen}, respectively. The data for both XIS FI detectors are co-added to form a single spectrum. Due to strong divergence in the low-energy responses, and the resulting strong data/model residuals, we discard all XIS data below 0.5 keV. We also discard data above 10 keV. Using the \ion{Mn}{i} K$\alpha_1$ and K$\alpha_2$ lines in the calibration spectrum, we measure a FWHM energy resolution of 149 eV at 5.9 keV for the present observation. The HXD/PIN data are also reduced following the {\em Suzaku} guidelines. The data are binned to a minimum of 20 counts per bin. We fit the data from 10--30 keV with a powerlaw and report a flux of $\rm (1.8 \pm 0.2)\times 10^{-11}$~erg~s$^{-1}$~cm$^{-2}$.

\subsection{{\it Chandra}} \label{sec:chandra-acis}
Mrk 1018 was observed with \textit{Chandra} ACIS \citep{garmire2003, weisskopf2000} only once in its bright type 1 phase in 2010 (C1; obsID 12868) for 23 ks. The data were reduced with the CIAO software using the CIAO 4.16.0 \texttt{'chandra\_repro'} script and the CALDB 4.11.0 set of calibration files. We filtered the data to eliminate bad grades and cosmic-rays.
Due to the high flux of the source, this spectrum was distorted due to pileup despite the fact that the observation used a 1/8 subarray that reduced pileup severity.
The CIAO \texttt{'pilep\_map'} script revealed that roughly $40$\% of the photons within the inner arcsecond of the point source were affected by pileup.
Therefore, we extracted the spectrum from the `readout streak', which comprise of photons that are detected while the data are being read-out (since the ACIS detector is shutterless). 
The short readout time results in these photons being unaffected by pileup, therefore providing an un-warped spectrum of the bright point source.
We used the CIAO \texttt{'dmextract'} function to extract the spectrum within the regions of the readout streak, and the \texttt{`mkacisrmf'} and \texttt{`mkarf'} tools to make the response files at the location of the point source.
The background spectra were extracted from regions adjacent to either side to the readout streak.
The exposure time was calculated by the integrated readout time of the observation across the streak length.
The 0.3--7 keV spectrum was used for the spectral modeling in this case.

All \textit{Chandra} spectra from the faint type 1.9 phase (C2-C8) are not affected by pile-up
We first reprocessed all the data using the \texttt{chandra\_repro} script. We then visually selected a circular source aperture with a radius of $\sim$3-4$''$ and an annular background region from the counts image, before extracting a sky-subtracted spectrum with the \texttt{specextract} tool using the source and background regions selected above. The extracted spectrum was then grouped using the \texttt{grppha} tool so that each bin contains a minimum of 10 to 20 counts.

\subsection{{\it Swift}} \label{sec:swift}
We use the {\it Swift} XRT and UVOT \citep{gehrels2004, burrows2005} observations  processed by the Space Science Data Center (SSDC)\footnote{\href{https://swift.asdc.asi.it/}{https://swift.asdc.asi.it/}} with the following ObsIds: 00035166001, 00035776001, 00049654001, 00049654002, and 00049654004 (hereafter Sw1, Sw2, Sw3, Sw4, Sw5) to estimate total X-ray flux, primarily to improve the sampling density of our flux light-curves in the pre-shutdown (<2016) phase. The pre-shutdown Sw1 and Sw2 XRT observations are piled up. We extract the source spectrum from annular regions with inner and outer radii 3$\arcsec$ and 20$\arcsec$, respectively, for Sw1 and Sw2 to reduce the effect of pile-up. For other observations, source spectra were reduced from a circular region of 20$\arcsec$. For all datasets, the background spectra were extracted from annular source-free regions with inner and outer radii 40$\arcsec$ and 80$\arcsec$).

Furthermore, we use the online service provided by  SSDC to generate the UVOT filter magnitudes and fluxes. Counts were extracted from a circular source of radius 5$\arcsec$ and an annular background region spanning radii 22$\farcs$4 -- 38$\farcs$2. The standard stars Star-1 and Star-3 
\citep[see][]{brogan2023} located within 1.2$\arcmin$ of Mrk 1018 were found to exhibit excess variability not consistent with their flux uncertainties, consequently requiring a flux correction. Here, we use a non-parametric approach for deriving the flux corrections, as described in Appendix \ref{apdx:lc-corr}. We de-redden these fluxes using the python package \texttt{extinction}\footnote{\href{https://extinction.readthedocs.io/en/latest/}{https://extinction.readthedocs.io/en/latest/}} before plotting and analysis, using $E(B-V)=0.0272$ obtained from the python package \texttt{sfdmap}\footnote{\href{https://pypi.org/project/sfdmap/}{https://pypi.org/project/sfdmap/}} \citep{schlegel1998}. For Sw2, we further use the sky images corresponding to each of the UVOT filters to generate \texttt{xspec} readable .pha files using \texttt{uvot2pha} for broadband spectral fitting.

\section{Data analysis} \label{sec:data-analysis}
We first analyze the high signal/noise X-ray spectra from \textit{XMM-Newton} and \textit{Suzaku} to establish the best broadband continuum X-ray spectral model in Sect.~\ref{sec:xmm-suzaku-analysis}. We then apply the best-fitting continuum model to the lower signal/noise {\it Swift} bright state and the {\it Chandra} X-ray datasets.

\subsection{{\it XMM-Newton} and {\it Suzaku} X-ray spectral analysis} \label{sec:xmm-suzaku-analysis}

For all \textit{XMM-Newton} observations, we jointly fit the pn0, pn14, MOS1, and MOS2 spectra, in the 0.3--10 keV for pn0, 0.5--8.0 keV for pn14, and 0.4--8.0 keV bands for MOS1 and MOS2. We allowed a relative constant factor to vary between all spectra except for pn pattern 0, where it is frozen at 1. Similarly, for the {\it Suzaku} spectra we kept the relative normalization variable for XIS1 and XIS3 free with that for XIS0 frozen at 1, and the best-fitting values were always close to 1.

For data fitting, we use Bayesian X-ray analysis \citep{skilling2004, feroz2009, buchner2014} to estimate parameter posteriors, and search for model preferences in \texttt{XSPEC} \citep{arnaud1996}. We use the C-statistic \citep[\texttt{cstat};][]{cash1979} for parameter optimization. Galactic absorption was frozen at $N_{\rm H} = 2.67 \times 10^{20}$~cm$^{-2}$ \citep{willingale13}, modeled using the \texttt{XSPEC} model \texttt{TBabs} \citep{wilms2000}. Cosmic abundances of \citet{wilms2000} and the photoelectric absorption cross section provided by \citet{verner1996} were used. All {\it XMM-Newton} \& {\it Suzaku} datasets were grouped to a minimum of 20 counts per bin. We first fit the spectra with simple models and progressively increase model complexity finally invoking a physically-motivated model:

\begin{itemize}
\item[1.] We initially fit a single power law (including Galactic absorption) model, \texttt{M1=TBabs*zpowerlw}, to all {\it XMM-Newton} and {\it Suzaku} datasets. This model did not return a good fit for any spectrum, (e.g. Fig. \ref{fig:residuals}) indicating that the X-ray spectrum contains multiple spectral components.

\item[2.] Fitting the datasets with a phenomenological double-power-law, \texttt{M2=TBabs(zpowerlw(1)+zpowerlw(2))}, diminishes the residuals (Fig.~\ref{fig:residuals} and Table~\ref{tab:xmm-suz-parameters}) significantly. For all observations, the photon indices for the steeper, soft-band power law ($\Gamma_1$), and the flatter, hard-band power law ($\Gamma_2$) remained constrained {\bf to} $1.96<\Gamma_1<3.14$ and $0.55<\Gamma_2<1.64$, respectively, with $\Gamma_2$ progressively becoming lower in time across our campaign. 
The steep photon indices of the soft-band power law indicate the presence of a soft X-ray excess in the spectrum, and the flat photon indices of the hard-band power law that are atypical of hot-Comptonized X-ray spectrum indicate the presence of a Compton reflection component above $\sim$ 6~keV, justifying the usage of the physically motivated soft-excess and a reflection model.
With respect to M1, the Bayesian evidence improved for M2 with $(\Delta \log Z)_{21} > 21.2$\footnote{$(\Delta \log Z)_{\rm ij} = \log Z_{\rm M_i} - \log Z_{\rm M_j}>$2 implies `$\rm M_i$' is a better model \citep{buchner2014}} in all six cases. 

\item[3.] We finally fit the datasets with a physically-motivated model, taking into account the warm Comptonization soft X-ray excess \citep{boissay2016} seen in the 0.3--2~keV band using \texttt{compTT} \citep{titarchuk1994} and distant Compton reflection from the torus using \texttt{uxclumpy} \citep{buchner2019}, alongside the primary hot-corona powerlaw: \texttt{M3 = TBabs*(compTT + zpowerlw + uxclumpy\_reflect)}.
We keep the following parameters frozen in this model as they will not be constrained by the data: \texttt{compTT} seed photon temperature $k_{\rm B}T_{\rm seed}$=0.01~keV, \texttt{uxclumpy} line of sight $N_{\rm H}$ to $1\times10^{20}$~cm$^{-2}$, angular cloud distribution parameter $\sigma_{0} = 84^{\circ}$ (the highest allowed value), \texttt{uxclumpy} Compton-thick ring covering factor $C_{\rm ring}$ to zero, and \texttt{uxclumpy} inclination $\theta_{i}$ to 45$^{\circ}$, since the source is neither obscured nor Compton-thick reflection-dominated \citep{husemann2016, brogan2023, veronese2024}. \texttt{M3} is the best-fitting model for SUZ, XMM3, XMM4, and XMM5, as indicated by values of $\log Z$ (Table \ref{tab:xmm-suz-parameters}; 3.6 $<$ (${\Delta}$log$Z$)$_{\rm 32}$ $<$ 22.4). For the XMM1 and XMM2 datasets, the phenomenological model M2 is marginally better ($(\Delta \log Z)_{\rm 32}>-3.93$, with similar values of test statistic $C$/dof for \texttt{M2} and \texttt{M3}), but does not rule out the physically motivated \texttt{M3} model in general. For XMM5, the soft excess was not significantly detected. A model consisting only of \texttt{TBabs*(zpowerlw + uxclumpy\_reflect)} provided a good fit, and inclusion of a \texttt{compTT} component, with $k_{\rm B}T_{\rm e}$ and $\tau$ frozen to the best-fitting values in XMM4, provided no significant improvement (${\Delta}$log$Z$ only $-$1.7). We find an upper limit to the 0.3--2.0 keV flux of the \texttt{compTT} component of $1.1 \times10^{-14}$ erg cm$^{-2}$ s$^{-1}$. We plot the best-fittng \texttt{M3} models for {\it XMM-Newton} and {\it Suzaku} in Fig.~\ref{fig:xmm-suz-spectra}.
We also test whether the built-in Fe K$\alpha$ line of the reflection model accounts for the total flux of the line in the data since it is self-consistently calculated using the XARS radiative transfer code code at a fixed (solar) abundance \citep{buchner2019}. Subsequently, we perform an analysis with an extra \texttt{zgauss} component added to \texttt{M3}
keeping all parameters frozen to the best-fit values from \texttt{M3} except the normalization parameters of the torus reflection component ($A_{\rm REFL}$) and \texttt{zgauss}. For the \texttt{zgauss} component, the rest wavelength and the width ($\sigma$) were frozen at 6.4~keV and 0.05~keV, respectively. The results show that $A_{\rm REFL}$ remains consistent with the \texttt{M3} fit, and the \texttt{zgauss} normalization is consistent with zero, suggesting that no additional Gaussian component is required, ruling out both significantly different non-solar abundances and significant extra emission from Compton-thin gas.
\end{itemize}

\begin{figure*}
    \centering
    \includegraphics[scale=0.68]{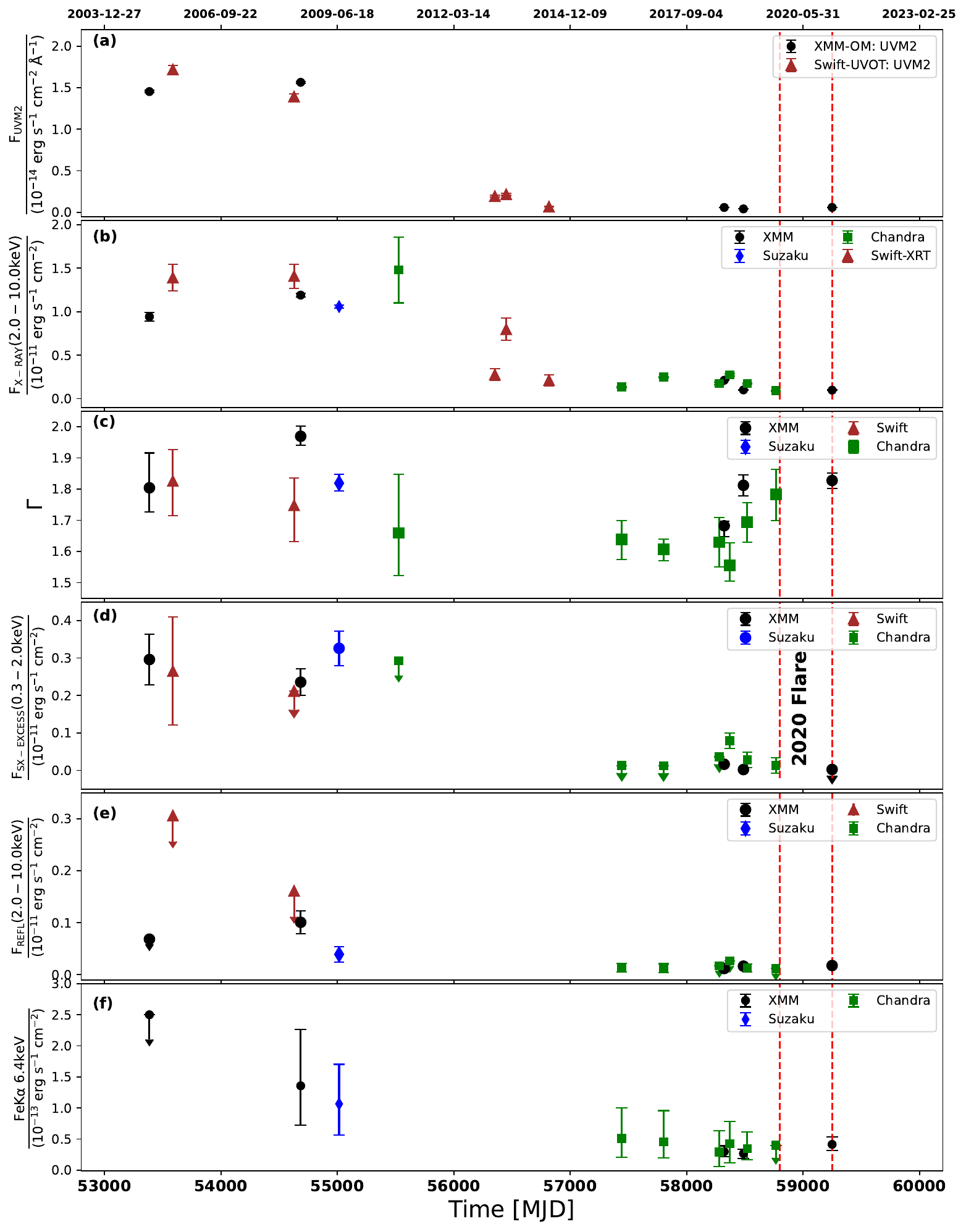}
    \caption{Overview of UV and X-ray spectral components.  
    (a) de-reddened and host-galaxy subtracted UVM2 flux from combined {\it XMM-Newton} OM and {\it Swift} UVOT data.
    (b) integrated total hard X-ray flux.
    (c) photon index ($\Gamma$) of the hot-corona power law (\texttt{zpowerlw }in \texttt{M3}; Sect.~\ref{sec:xmm-suzaku-analysis}).
    (d) integrated flux of the soft excess flux component (0.3--2.0~keV)
    (e) integrated flux of the reflection component (2.0--10.0~keV)
    (f) integrated Fe~K$\alpha$ emission line flux.
    Black-circular points: {\it XMM-Newton}; brown triangular points: {\it Swift}; blue diamond point {\it Suzaku}; and green square points: {\it Chandra}. The red vertical lines mark the beginning and end of the 2020 outburst in the optical band \citep{brogan2023}.}
    \label{fig:lc_all}
\end{figure*}

Henceforth, we adopt the \texttt{M3} model fits for estimation of the continuum properties of the X-ray spectra. We plot the flux variability of all \texttt{M3} model components in Fig.~\ref{fig:lc_all}.

\begin{table*}
\caption{Summary of ultraviolet (UVM2 band) flux density, total 2--10~keV X-ray flux, hot-corona photon index ($\Gamma$, from model \texttt{M3}) and Fe~K$\alpha$ emission line properties (Sect.~\ref{sec:iron-line-analysis}).}\label{tab:fluxes-photon-indices}
\centering
    \begin{tabular}{ccccccccccccccc}
    \hline
    \hline
    Time & Inst. & UV &                            \multicolumn{2}{c}{X-ray continuum} &   $\log \lambda_{\rm Edd}$~$^c$     & \multicolumn{3}{c}{Fe K$\alpha$ properties} \\
    \hline
         &      &  F$_{\rm UVM2}$ $^{a}$       & $\Gamma$  &   F(2--10~keV)$^b$  &  & Flux$^d$ & Eq-Width & $\sigma_{\rm Fe~K\alpha}$ \\ 
    (yyyy-mm-dd) & & & & & &   & (keV) & (eV)\\
    \hline
   \multicolumn{8}{c}{\textit{bright type 1 phase}} \\
   \hline
   2005-01-15 & XMM1    &$1.45\pm0.01$   &$1.80^{+0.11}_{-0.08}$& $9.3 \pm 0.6$  & $-1.17 \pm 0.06$ & $<2.50$ & $<0.180$       & 50 \\
   2005-08-05 & Sw1 &$1.72 \pm 0.05$ &$1.89 \pm 0.07$    &$14.1 \pm 0.8$  & $-0.98 \pm 0.04$ & --      & --             & -- \\
   2008-06-11 & Sw2 & $1.39 \pm 0.03$ & $1.76 \pm 0.09$  & $12.7 \pm 2.0$ & $-0.98 \pm 0.05$ & --      & --             & -- \\
   2008-08-07 & XMM2    & $1.56 \pm 0.01$ & $1.97 \pm 0.03$  & $11.8 \pm 0.2$ & $-1.06 \pm 0.01$ &$1.38^{+1.20 }_{-0.74 }$& $0.07 \pm 0.05$ & $135^{+360}_{-110}$\\
   2009-07-03 & Suz     & --  & $1.86 \pm 0.02$ & $10.6 \pm 0.2$ & $-1.11 \pm 0.01$ &$1.07^{+0.64 }_{-0.5 }$ & $0.11^{+0.03}_{-0.05}$ &$195^{+145}_{-101}$\\
   2010-11-27 & C1      & --              & $1.68^{+0.12}_{-0.13}$ & $12.4^{+1.8}_{-2.3}$ & $-0.96 \pm 0.12$  & --                 & --                        & --\\
    \hline
   \multicolumn{8}{c}{\textit{transition phase}} \\
   \hline
   2013-03-01 & Sw3 & $0.19 \pm 0.01$  & -- & $2.8 \pm 0.7$ & $-1.72 \pm 0.12$ & -- & -- & -- \\
   2013-06-07 & Sw4 & $0.22 \pm 0.01$  & -- & $8.0 \pm 1.3$ & $-1.24 \pm 0.08$ & -- & -- & -- \\
   2014-06-09 & Sw5 & $0.07 \pm 0.01$& -- & $2.1 \pm 0.6$ & $-1.85 \pm 0.13$ & -- & -- & -- \\
   \hline
   \multicolumn{8}{c}{\textit{faint type 1.9 phase}} \\
   \hline
   2016-02-25 & C2           & --              & $1.64 \pm 0.06$        &$1.20 \pm 0.09$ & $-2.05 \pm 0.03$ & $0.51 ^{+0.50 }_{-0.31 }$ & $0.53^{+0.16}_{-0.15}$ & $140^{+173}_{-87}$\\
   2017-02-17 & C3           & --              & $1.61^{+0.03}_{-0.04}$ &$2.3 \pm 0.1$   & $-1.78 \pm 0.03$ & $0.43 ^{+0.63 }_{-0.31 }$ & $0.09 \pm 0.08$ & $100^{+110}_{-70}$\\
   2018-06-12 & C4           & --              & $1.63 \pm 0.08$        & $1.7 \pm 0.1$  & $-1.93 \pm 0.02$ & $0.29 ^{+0.31 }_{-0.23 }$ & $0.19 \pm 0.11$  & 50\\
   2018-07-23 & XMM3       & $0.06 \pm 0.01$ & $1.68^{+0.01}_{-0.03}$ & $2.1 \pm 0.04$  & $-1.84\pm 0.01$  & $0.3 ^{+0.10 }_{-0.08 }$ &  $0.17 \pm 0.06$& $72^{+65}_{-60}$\\
   2018-09-09 & C5           & --              & $1.56^{+0.07}_{-0.05}$ & $2.7 \pm 0.13$  & $-1.73 \pm 0.02$ & $0.38 ^{+0.35 }_{-0.28 }$& $0.165^{+0.193}_{-0.163}$ & 50 \\
   2019-01-04 & XMM4       & $0.04 \pm 0.01$ & $1.82^{+0.02}_{-0.04}$ & $0.84 \pm 0.03$ & $-2.17 \pm 0.01$ & $0.27^{+0.08 }_{-0.09}$  & $0.27 \pm 0.10$          & $<80$\\
   2019-02-(06+07) & C6+C7 & --            & $1.69 \pm 0.06$        & $1.6 \pm 0.1$	  & $-1.93 \pm 0.02$ & $0.25 ^{+0.31 }_{-0.22 }$& $0.168^{+0.159}_{-0.154}$ & 50 \\
   2019-10-10      & C8$^{\dag}$& --           & $1.78 \pm 0.08$        & $0.86 \pm 0.07$ & $-2.22 \pm 0.03$ & <0.37 & <0.41    & 50 \\
   2021-02-04 & XMM5 & $0.06 \pm 0.01$ & $1.84\pm 0.02$ & $0.80 \pm 0.02$ & $-2.18 \pm 0.01$ & $0.43 ^{+ 0.12 }_{- 0.12 }$& $0.37 \pm 0.11$ &$110^{+48}_{-42}$ \\
    \hline
    \hline
    \end{tabular}
    \tablefoot{All \textit{XMM-Newton} estimates are based on joint fitting of the pn0, pn14, MOS1, and MOS2 datasets. The {\it Suzaku} estimate is based the joint XIS0, XIS1, and XIS3 datasets. \\
    \tablefoottext{$a$}{Flux density in units of $10^{-14}$~erg~cm$^{-2}$~s$^{-1}$~$\AA^{-1}$, host galaxy subtracted\\}
    \tablefoottext{$b$}{Flux in units of $10^{-12}$ erg~cm$^{-2}$~s$^{-1}$ \\}
    \tablefoottext{$c$}{$\lambda_{\rm Edd}$ calculated using bolometric correction from \citet{duras2020} \\}
    \tablefoottext{$d$}{Integrated line flux in units of $10^{-13}$ erg~cm$^{-2}$~s$^{-1}$.}}
\end{table*}

\begin{table*}
    \caption{\texttt{agnsed} fits to broadband SEDs (Sect.~\ref{sec:broad-band})} \label{tab:agnsed}
    \centering
    \begin{tabular}{lcccccc}
    \hline
    \hline
    Parameters           & XMM1                   & XMM2                    & XMM3                   & XMM4                   & XMM5  \\

    \hline
    $\log \lambda_{\rm Edd}$~$^a$ & $-1.37^{+0.02}_{-0.01}$ & $-1.32^{+0.01}_{-0.01}$ & $-2.17^{+0.01}_{-0.02}$& $-2.52^{+0.01}_{-0.01}$ & $-2.53 \pm 0.01$ \\
    $k_{\rm B}T_{\rm wc}$ (keV)& $0.17^{+0.03}_{-0.02}$& $0.16^{+0.02}_{-0.01}$  & 0.2$^{f}$ & 0.17$^{f}$    &  - \\
    
    $\Gamma_{\rm WC}$ & $2.07^{+0.15}_{-0.07}$  & $2.19^{+0.12}_{-0.09}$  & $1.87^{+0.19}_{-0.16}$ & $2.56^{+0.17}_{-0.09}$ &  -  \\
    $R_{\rm hot}$  ($R_{\rm g}$) & $25.7 \pm 1.0$& $28.5^{+0.5}_{-0.7}$ & $73.7^{+1.5}_{-1.9}$ & $43.4^{+3.2}_{-2.1}$ &  $47.0 \pm 1.0$ \\
    $R_{\rm warm}$ ($R_{\rm g}$) & $35.5^{+3.0}_{-2.0}$ & $42.6^{+4.7}_{-3.2}$ & $84.8^{+5.3}_{-3.4}$ & 84.8$^{f}$  & -\\
    $\alpha_{\rm OX}$ & $1.33 \pm 0.01$ & $1.28 \pm 0.01$ & $1.08 \pm 0.01$ & $1.14 \pm 0.01$ & $1.21 \pm 0.01$\\
    $\left[ \frac{F_{\rm AGN}}{F_{\rm AGN} + F_{\rm gal}} \right]_{\rm UVM2}$ & 0.99& 0.99& 0.89 & 0.86 & 0.90\\
    $L_{\rm UV}$~$^{b,d}$ & $2.41 \pm 0.05$  & $2.48^{+0.04}_{-0.03}$ & $0.12 \pm 0.01$ & $0.086^{+0.004}_{-0.003}$ & $0.102\pm0.002$\\
    $L_{\rm X-ray}$~$^{c,d}$ & $1.11 \pm 0.06$ & $1.59 \pm 0.04$ & $0.17 \pm 0.01$ & $0.081 \pm 0.003$ & $0.078 \pm 0.00$\\
    $\frac{L_{\rm X-ray, 2-10~keV}}{L_{\rm UV, 2-15~eV}}$ & $0.46 \pm 0.03$ & $0.64 \pm 0.02$ & $1.45 \pm 0.08$ & $0.94 \pm 0.05$ & $0.76\pm0.02$ \\

    \hline
    \hline
    \end{tabular}
    \tablefoot{The \texttt{agnsed} fits are based on different sets of available filters as listed in Table \ref{tab:observation-list}.\\ 
    \tablefoottext{$a$}{Accretion rate relative to Eddington. }\\
    \tablefoottext{$b$}{integrated in the 2--15~eV band}\\
    \tablefoottext{$c$}{integrated in the 2--10~keV band}\\
    \tablefoottext{$d$}{in units of $10^{44}$~erg~s$^{-1}$}\\
    \tablefoottext{$f$}{indicates frozen parameter}}
\end{table*}

\begin{figure*}
    \centering
    \includegraphics[scale=0.45]{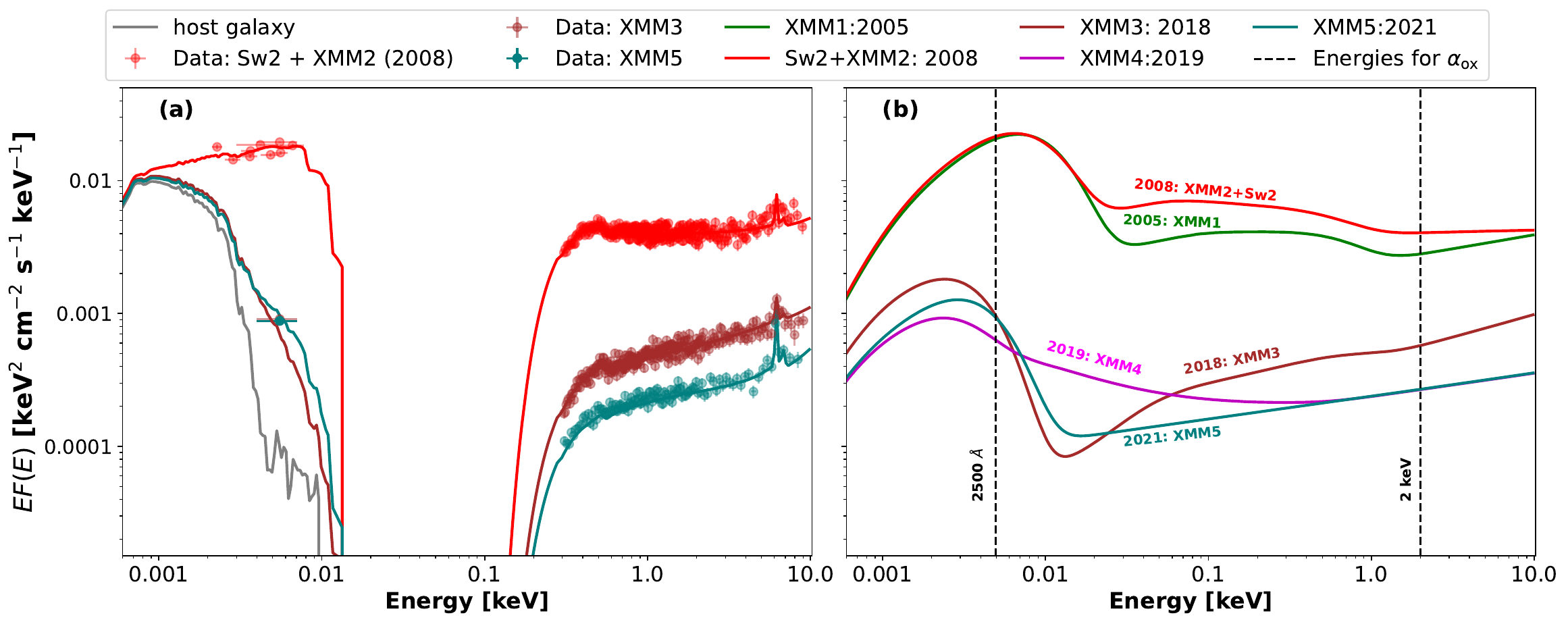}
    \caption{\texttt{agnsed} broadband SED fits.
    (a) The absorbed total broadband model. The circular markers indicate {\it XMM-Newton} EPIC-pn+OM datasets for the XMM3, and XMM5 observations and {\it XMM-Newton} EPIC-pn + OM + {\it Swift}-UVOT datasets for XMM2. The solid lines of corresponding color represents the best-fit total models.
    The XMM1 and XMM4 data points are not plotted for clarity. For XMM3 and XMM5 the OM photometric points almost overlap with each other since they have almost equal fluxes.
    (b) The corresponding unabsorbed \texttt{agnsed} models. They indicate the significant changes occurring in the intrinsic spectrum, within both the bright and faint flux states. The dashed lines indicate energies used to calculate the optical-X-ray spectral index $\alpha_{\rm OX}$.}
    \label{fig:agnsed}
\end{figure*}

\subsection{{\it Chandra} X-ray spectral analysis} \label{sec:chandra-analysis}
We apply the inferred best-fitting physically-motivated continuum model, \texttt{M3}, to all \textit{Chandra} datasets. We ignore all ACIS data below 0.7 keV since the drop in effective area yields too few counts for reliable constraints on soft excess model parameters, and we fit up to 8.0 keV. We fix the \texttt{compTT} temperature ($k_{\rm B}T$) at 0.17~keV, the optical depth ($\tau$) at 20, while keeping the normalization ($A_{\rm compTT}$) free. Observations C6 and C7 were taken within a day, so we fit both datasets simultaneously, allowing only a variable normalization factor between them. The average flux of C6 and C7 is reported in Table \ref{tab:fluxes-photon-indices} and plotted in all flux light curves presented in this paper.

\subsection{{\it Swift} X-ray spectral analysis} \label{sec:swift-analysis}
For the {\it Swift}-XRT datasets we used data in the 0.3--8.0 keV range. We apply the inferred best-fitting physically-motivated continuum model, \texttt{M3}, to the bright state Sw1 and Sw2 datasets to infer spectral properties (e.g. $\Gamma$) and individual spectral component fluxes. The soft excess optical depth was kept frozen at $\tau$=20 for both cases. Best-fitting values for soft excess temperature were  $0.12^{+0.06}_{-0.05}$~keV for Sw1 and $<$0.16~keV for Sw2.

The low count (<300) intermediate state spectra: Sw3, Sw4, and Sw5, were fit with only a power law to evaluate total model flux.

\subsection{The Fe K$\alpha$ line} \label{sec:iron-line-analysis}
The isolated Fe K$\alpha$ line flux cannot be derived from the \texttt{M3} model fitting, as the publicly available UXCLUMPY reflection model does not provide a separate component for the Fe~K$\alpha$ emission line. Therefore, we adopt a phenomenological approach, applying a model \texttt{M$_{\rm cont}$ + TBabs*$\Sigma_{\rm i}$zgauss$_{\rm i}$}, where $M_{\rm cont}$ is an underlying continuum and \texttt{zgauss} are the emission line component(s). For neutral or moderately-ionized gas, Fe K$\alpha$ line emission is accompanied by Fe K$\beta$ (7.06 keV rest frame). Due to the close energy proximity of these lines, we account for the potential flux contribution from the Fe K$\beta$ line even though it is unresolved due to insufficient counts. We thus add a Gaussian component with width $\sigma$ tied to that for K$\alpha$ and normalization tied to 13\% \citep{palmeri2003} that of the K$\alpha$ line.

For the high-count 
spectra, XMM2, SUZ, XMM3, XMM4, and XMM5, we adopt
 \texttt{M2} for the continuum component \texttt{M$_{\rm cont}$} 
(Sect.~\ref{sec:xmm-suzaku-analysis}). Given the substantial proportional uncertainties (up to 32\%) in the hard photon indices (Table \ref{tab:xmm-suz-parameters}), we freeze $\Gamma_1$ and $\Gamma_2$ to the values derived from the continuum fit (Sect.~\ref{sec:xmm-suzaku-analysis}). Meanwhile, the Gaussian width $\sigma_{\rm Fe~K\alpha}$, its normalization, and the normalization values of the two power laws $A_{1}$ and $A_{\rm 2}$) are allowed to vary. We keep the Gaussian line centroid frozen at 6.4~keV (rest frame) for all \textit{XMM-Newton} datasets though not for the {\it Suzaku} dataset, where keeping line centroid free returned a better fit. 

For the {\it Chandra} datasets, we adopt a power law as the underlying continuum. In all fits except for C2 and C6+C7, we keep the line centroids and widths frozen to 50~eV. For C2 we keep both width and centroid free, and for C6+C7 we keep only the line centroid free (Table \ref{tab:fluxes-photon-indices}).

We use the Monte Carlo simulations \texttt{simftest} to calculate the detection probability of the Fe~K$\alpha$ line. For all {\it XMM-Newton} and {\it Suzaku} observations, we obtain a detection probability $>$99\% case except for the XMM1 spectrum. The Fe~K$\alpha$ line was detected at high confidence ($>95$\%, with significance reaching $>99\%$ confidence in C2 and C6+C7) in all {\it Chandra} observations except C1 and C8.
 
We list the Fe K$\alpha$ line parameters, including equivalent width relative to the local continuum, in Table \ref{tab:fluxes-photon-indices} and plot the Fe K$\alpha$ flux in Fig.~\ref{fig:lc_all}.

\begin{figure}
    \centering
    \includegraphics[scale=0.54]{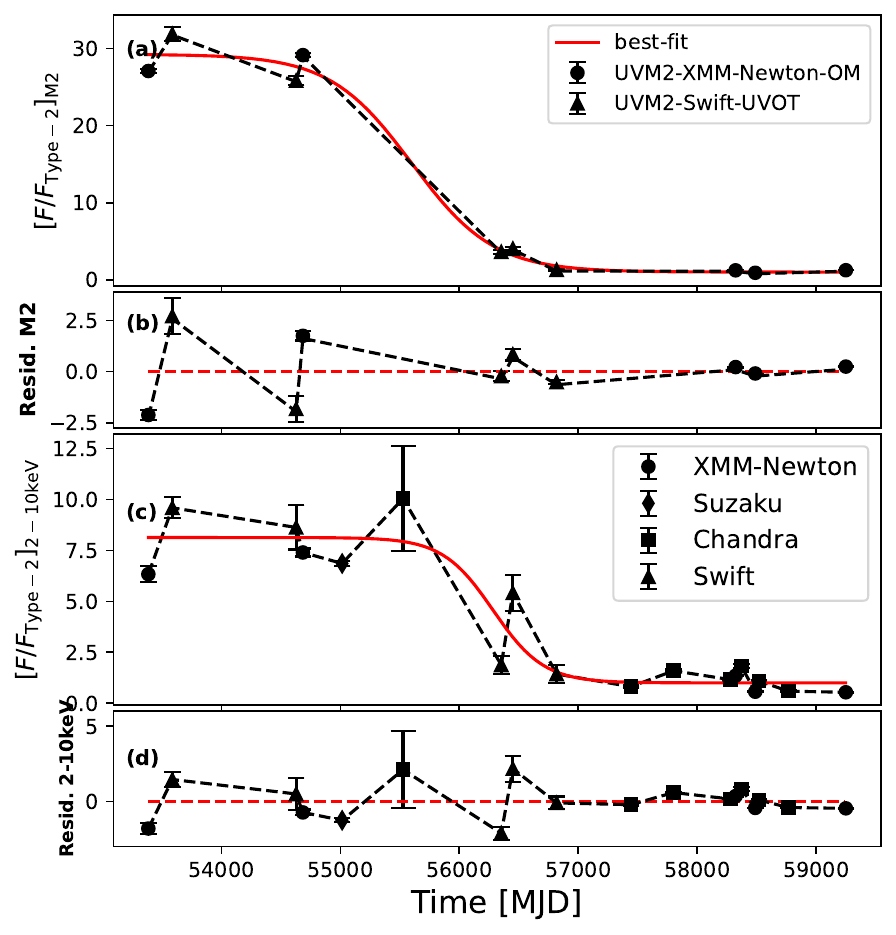}
\caption{(a) UVM2 light-curve data and best-fit MRS function normalized by the faint state flux (b) corresponding UVM2 residual (c) 2-10~keV X-ray light-curve data and best-fit normalized by the faint state flux(d) corresponding X-ray residual. The circular, diamond, triangular, and the square markers represent {\it XMM-Newton}, {\it Suzaku}, {\it Swift} and {\it Chandra} datasets, respectively. The red line indicates the best-fit MRS model.}\label{fig:lcfit}
\end{figure}

\subsection{Broadband SED analysis} \label{sec:broad-band}
The multi-band behavior of the Mrk 1018 is constrained by the joint fitting of the optical, UV, and X-ray observations. We fit all the high signal/noise {\it XMM-Newton} EPIC-pn and OM datasets using a broadband spectral model; we also included Sw2 in the fit to XMM2. The broadband spectral model consists of a host galaxy template \citep{mannucci2001}, AGN disk+corona emission from \texttt{agnsed} \citep{done18}, and distant reflection modeled with \texttt{uxclumpy}, mathematically written as:\\
\texttt{$M_{\rm agnsed}$=redden*TBabs*(host\_galaxy + agnsed + uxclumpy\_reflect)}. \\
The \texttt{redden} and \texttt{TBabs} components account for the Galactic absorption in the optical/UV and X-ray bands, respectively. For Mrk 1018, $E(B-V)$=0.0272 \citep{schlegel1998}, and the comoving-distance is $d_{\rm 0}=182$~Mpc. Furthermore, we always keep the following \texttt{agnsed} parameters: hot corona temperature $k_{\rm B}T_{\rm e}=200$~keV, corona height($h_{\rm x}$) = $10~R_{\rm g}$, outer radius $R_{\rm out}$ set to the self-gravity radius \citep[][and references therein]{done18}, black hole mass $M_{\rm BH}=10^{7.9}M_{\rm \odot}$ \citep{mcelroy2016}, $\theta_{\rm i}=45^{\circ}$, and black hole spin $a$=0. The model setup considered X-ray reprocessing in the accretion disk.
The broadband spectra of Mrk 1018 contain a significant amount of host galaxy flux relative to the AGN, as indicated by the optical images and spectra in \cite{mcelroy2016} and \cite{brogan2023}. The \textit{XMM-Newton} OM observations did not take data at V and B bands, necessary for optimal constraints on the host galaxy contribution to the SED. Therefore, we keep the host galaxy normalization frozen at the host galaxy level obtained from the optical bands via joint spectral-fitting of the multiwavelength XMM2 and Sw2-UVOT observations from 2008. The host galaxy fraction is just ($F_{\rm host~galaxy}/F_{\rm tot}$) = $4.78\times 10^{-3}$ in the UVM2 (4.25--6.81~eV) band.
Consequently, the host galaxy has a flux density of $F_{\rm gal, UVM2}=6.98 \times 10^{-17}$~ergs~cm$^{-2}$~s$^{-1}$~$\AA^{-1}$ in the UVM2 band.
For XMM5, we demonstrated in Sect.~\ref{sec:xmm-suzaku-analysis} that the warm coronal component is not robustly detected. Thus, for fitting the XMM5 data in the \texttt{agnsed} framework, we do not include such a component.
We report the best-fit values in Table \ref{tab:agnsed}, 
plot the best-fit spectra in Fig.~\ref{fig:agnsed}, and interpret the results in Sect.~\ref{sec:multiwavelength}.

As a caveat, we bear in mind that the application of the AGNSED model in the faint state is approximate, as the faint state is potentially dominated by an ADAF. Thus, it can have additional radiative contributions from synchrotron, cyclotron, and bremsstrahlung \citep{narayan1995} processes. 
This additional emission can modify the broad band spectral shape, however, an alternative broad band model that incorporates these radiative processes added to the baseline radiation from the AGNSED model, is currently not available.
This additional radiative contribution if present can potentially influence our estimates of the model-dependent parameters like $R_{\rm hot}$, $R_{\rm warm}$, $\Gamma_{\rm warm}$, $kT_{\rm warm}$, etc. However, our estimate of the $\alpha_{\rm OX}$ and Eddington ratio are only mildly dependent on spectral modeling and will potentially remain unaffected.

\subsection{Continuum X-ray and UV flux variability modeling} \label{sec:flux_variability}
To visualize and quantify the variability of the long-term flux drop $f_{\rm long}(t)$, we fit the X-ray and UV data with a phenomenological `modified reverse sigmoid'(MRS) function:
\begin{equation}
    f_{\rm long}(t) = F_{\rm faint} \left( \frac{R-1}{1 + e^{(t-t_0)/t_{\rm sc}}} + 1  \right)
\end{equation}
The description of the parameters are the following:
\begin{itemize}
    \item $F_{\rm faint}$: average flux in the faint state
    \item $R$: ratio of the bright to the faint state flux
    \item $t_0$: defined such that $f_{\rm long}(t_{0}) = \frac{F_{\rm bright}+F_{\rm faint}}{2} = F_{\rm faint} \left( \frac{R+1}{2} \right)$
    \item $t_{\rm sc}$: timescale quantifying how fast the flux drops; smaller value implies faster drop
\end{itemize}
The average long-term flux drop is well captured by the analytical model (Fig.~\ref{fig:lcfit}). Best-fitting values of $R$ were 7.9$\pm$0.5 and 28$\pm$2 for hard X-rays (2--10~keV) and UV, respectively; $t_{0}$ was found to be MJD $56228\pm$195 for X-rays and $55453 \pm 245$ for UV. Best-fitting values of $t_{\rm sc}$ were $274 \pm 187$~days for X-rays and $422 \pm 109$~days for UV, with their mean around 1 year.
The quantity $\left[\frac{R}{t_{sc}}\right]$ quantifies the rate of flux drop; best-fitting values for the X-ray and the UV light-curves were $0.03^{+0.06}_{-0.01}$ day$^{-1}$ and $0.07^{+0.02}_{-0.01}$ day$^{-1}$, respectively. We note however, as a caveat, that sparse sampling between 2014 and 2016 can impact our estimates of $t_0$, $t_{\rm sc}$, and $\left[\frac{R}{t_{sc}}\right]$. Additionally, uneven sampling between the X-ray and UV bands, means that we cannot exclude that both bands dropped simultaneously with the same slope. An additional impact is the short-term stochastic variability observed at all flux states of the source.

We search for lags or leads between the 2--10~keV and UVM2 flux light curves, using the Interpolated Correlated Function (ICF; \citealt{white1994}; based on \citealt{gaskell86}). To estimate uncertainties, we consider random subset selection and flux randomization following \citet{peterson98}. The pair of light curves are highly correlated but with no evidence for lags or leads in either case, with lags consistent with zero. For lags from 2--10 keV to UVM2, we find a maximum correlation coefficient $r_{\rm corr}$ = 0.962 at a delay of $\tau =-640 \pm 962$~days, i.e. consistent with zero lag.

\begin{table*}
    \centering
    \caption{Average flux of the UVM2 (5.32~eV) band, X-ray spectral components, and bolometric luminosity with there respective flux drop.}
    \begin{tabular}{@{}lccc@{}}
    \hline
    \hline
        Spectral component & \multicolumn{2}{c}{Average flux/Luminosity} & Factor$^{e}$ \\
        or Waveband & Bright & Faint& ($R_{\rm Component}$ or \\
       & & &  $R_{\rm Waveband}$)\\ 
        \midrule
        UVM2 (de-reddened) $^{a,d}$       & $1.54 \pm 0.03$ & $0.061 \pm 0.002$  & $25.3 \pm 0.6$\\
        $L_{\rm UV}$ $^{b,c,d}$ (0.002--0.015~keV) & $2.44 \pm 0.04$ & $0.104 \pm 0.005$  & $23.5 \pm 0.8$\\
        Hot power law (2--10~keV)               & $1.24 \pm 0.13$ & $0.16 \pm 0.01$  & $8.0 \pm 0.9$\\
        Soft-excess (0.3--2~keV)                & $0.27 \pm 0.05$ & $0.02 \pm 0.01$  & $12.0 \pm 4.0$\\
        Reflection (2--10~keV)                  & $0.069 \pm 0.013$&$ 0.016 \pm 0.036$ & $4.4 \pm 1.3$\\
        Ionizing flux Fe K$\alpha$ (7.1--10.0~keV) & $0.37\pm 0.05$   & $0.054 \pm 0.003$  & $6.8 \pm 1.1$\\
        Fe K$\alpha$                  & $0.012\pm 0.004$ & $0.004\pm 0.002$ & $3.4 \pm 2.5$\\
        $L_{\rm bol}$ $^b$ \citep{duras2020} & $9.20 \pm 0.90$ & $1.13 \pm 0.05$ & $8.2 \pm 0.9$\\
        Accretion rate ($\dot{M}$) $^f$ & $0.28\pm 0.03$ & $0.035\pm0.002$ & $8.2 \pm 0.9$\\
        
    \hline
    \hline
    \end{tabular}
    \tablefoot{ The bright type 1 and faint type 1.9 average flux is for time before MJD 56000 and after MJD 57000, respectively.\\
    \tablefoottext{$^a$}{in units of $10^{-14}$~erg~cm$^{-2}$~s$^{-1}$~$\AA^{-1}$}\\
    \tablefoottext{$^b$}{in units of $10^{44}$~erg~s$^{-1}$}\\
    \tablefoottext{$^c$}{integrated \texttt{agnsed} model luminosity}\\
    \tablefoottext{$^d$}{de-reddened and host-galaxy subtracted}\\
    \tablefoottext{$^f$}{in units of $M_{\rm \odot}$ yr$^{-1}$}\\
    \tablefoottext{$^e$}{ratio of the average fluxes}\\
    All X-ray fluxes are in units of $10^{-11}$~erg~cm$^{-2}$~s$^{-1}$}
    \label{tab:component-variability}
\end{table*}

\section{Results and discussion} \label{sec:discussion}

\subsection{Summary of main results} \label{disc:results}
Previous X-ray studies of Mrk~1018 have generally used rather simple models to quantify its complex X-ray spectrum and spectral behavior.
A single power law \citep{lyu2021}, double-power law \citep{brogan2023}, or/and a power law + Comptonization model \citep[\texttt{nthcomp}, \citealt{zdziarski1996} in][]{veronese2024} have been applied to characterize the X-ray continuum. 
Here, we fit all high signal/noise X-ray spectra with a physically-motivated three-component model that incorporates the primary hot coronal X-ray power law, a warm Comptonization soft X-ray excess, and a reflection component, including a narrow Fe K$\alpha$ line.

Our X-ray spectral decomposition demonstrated that the soft excess flux drops by a factor of $\sim$12 compared to bright-state across all observations, along with significant short term variability through the entire period of its detection until 2019. However, despite this strong flux drop, the parameters of the warm corona (optical depth $\tau$, electron temperature $k_{\rm B}T_{\rm e}$) do not evolve significantly through the entire period of its detection until 2019. Ultimately, the soft excess was not detected anymore in the high-count {\it XMM-Newton} spectrum in 2021.

Meanwhile, the the hot corona photon index $\Gamma$ demonstrates distinct trends: first decreasing across the bright-to-faint transition, and then increasing again after roughly 2018, when Mrk 1018 was in the faint state. These trends will be discussed below in more detail.

The UV-to-X-ray emission can be modeled with the thermal Comptonization model \texttt{agnsed}, and best-fitting X-ray parameters (e.g., $k_{\rm B}T_{\rm e}$) are generally consistent with the best-fitting \texttt{M3} models to the X-ray-only data. Similarly, \texttt{agnsed} does not return a soft-X-ray excess contribution for XMM5, consistent with the X-ray-only fitting (Fig. \ref{fig:agnsed}). We characterize the broadband spectral evolution using the UV-X-ray spectral index $\alpha_{\rm OX}$ \citep{tananbaum1979}. Values are listed in Table \ref{tab:agnsed}. $\alpha_{\rm OX}$ decreases from XMM1 to XMM3, but then increase through XMM4 and XMM5; a similar trend is also seen in the hot-corona powerlaw X-ray photon index ($\Gamma$).

\begin{figure}
    \centering
    \includegraphics[scale=0.55]{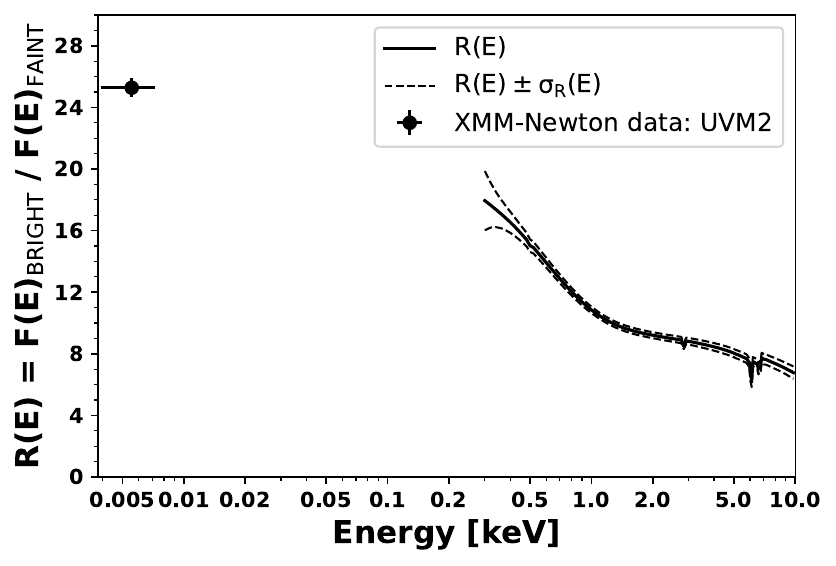}
    \caption{Energy resolved bright and faint state flux ratio. The solid black line is the ratio of the average X-ray spectra ($R(E)$) in the bright and the faint phase and the dashed lines demarcate the upper and lower error. The plot is based on the \texttt{M3} model \texttt{compTT + zpowerlw + uxclumpy-reflect} calculated for the high-count {\it XMM-Newton} and {\it Suzaku} spectra. The UVM2 flux ratio (diamond marker) is derived from {\it XMM-Newton} OM observations.}\label{fig:spectra_change_between_types}
\end{figure}

All emission components dropped in intensity during the spectral transition but by differing factors. To quantify these differences we examine the overall evolution in broadband spectral shape from the bright state to the faint state. We define a ratio spectrum, $R(E)$, to indicate the ratio of the flux density in the bright state to that in the faint state. $R(E)$ is determined from the average best-fitting model parameters for three spectra in each state, as detailed in Appendix \ref{sec:ratio-spec}. We plot $R(E)$ as a function of energy in Fig. \ref{fig:spectra_change_between_types}. The strongest contributors to evolution of the broadband spectral shape are the UV-band and the soft X-ray band. The average fluxes for each spectral component in the two states are listed in Table~\ref{tab:component-variability}.

Comparing the flux drops across the Fe K$\alpha$ line, the hard X-ray power law, the soft excess, and the UV luminosity (Table \ref{tab:component-variability}), defined as $R_{\rm Fe~K\alpha}$, $R_{\rm HX}$, $R_{\rm SX}$, and $R_{\rm UVM2}$, respectively, we note that $R_{\rm Fe~K\alpha}$ $<$ $R_{\rm HX}$ $<$ $R_{\rm SX}$ $<$ $R_{\rm UV}$, roughly consistent with the shape of the broadband $R(E)$ curve in Fig.~\ref{fig:spectra_change_between_types}. The results suggest that there is an overall broadband spectral hardening due to the CLAGN transition from the bright to faint state, which results from the interplay between different spectral components and bands. However, individual spectral components exhibit divergent behavior as described in the subsequent sections.

\subsection{The warm corona in Mrk 1018}\label{disc:warm-corona}
Mrk 1018 exhibits a drastic drop in the flux of the soft X-ray excess - a factor of 12 between the average bright and faint states (Table \ref{tab:component-variability}). In the context of the warm Comptonization model \citep[e.g.,][]{boissay2016}, however, there is no systematic change in the warm corona parameters through the entire period of its detection until 2019. We find that the temperature is always in the range 0.13~keV$<k_{\rm B}T_{\rm e}<$0.2~keV and opacity values are optically thick with $\tau>10$, which are consistent with studies by \cite{mehdipour2011, petrucci2013, ballantyne2024, palit2024}. 
One possible explanation for the decrease in soft excess until 2019 in the context of warm Comptonization is that due to the decrease in $\lambda_{\rm Edd}$, fewer optical-UV thermal seed photons from the accretion disk are incident on the warm corona.
We note that values of $F_{\rm SX,compTT}/F_{\rm tot}$, as listed in Table~\ref{tab:xmm-suz-parameters}, very roughly linearly track the corresponding values of $\lambda_{\rm Edd}$.
This scenario does not need to invoke any changes in the structure or properties of the warm corona during that process.

On the other hand, during the faint state, other physical processes seem to be active. From XMM3 (2018) to XMM4 (2019) and XMM5 (2021), the soft excess contribution decreased by at least a factor of $\sim$3. Since the soft excess component was not detected in the high signal/noise XMM5 data from 2021, its upper limits indicate only another decrease by a factor of $\sim$2 compared to XMM4. In the same time (between XMM3, XMM4, XMM5), the UV seed photon emission varied only by $\lesssim$30\% (Tables \ref{tab:fluxes-photon-indices}, \ref{tab:agnsed} and Fig. \ref{fig:agnsed}). Interestingly, XMM3 and XMM5 even have very similar UVM2 flux levels. 
Thus, it is unlikely that the further soft excess decrease below our detection limit (or its disappearance) after 2019 is driven solely by the variation in accretion disk seed photons. Since the seed photon production does not change significantly in the faint state, an alternate plausible explanation is a delayed (with respect to the time of CLAGN detection) structural change, disintegration, or disappearance of the warm corona (producing the soft excess). 
This notion is supported by the SEDs of samples of unobscured Seyferts studied by \citet{hagen2024} in the context of the geometry assumed in the \texttt{agnsed} model. The radiatively efficient inner flow along with the warm corona becomes absent for $\lambda_{\rm edd} \lesssim 0.02$.
This results in the negligible interception of cooler accretion disk seed photons in the faint state. The exact mechanism of warm corona disintegration or disappearance is still unknown. However, it is proposed that with the decrease in accretion rate, the flow transforms into a radiatively inefficient advection dominated accretion flow \citep[ADAF,][]{narayan1995, esin1997} and subsumes the radiatively efficient components of the accretion flow.

Multiple other sources also exhibit extreme soft excess variability tied to rapid accretion rate changes. Recent studies of the CLAGN NGC 1566 report the appearance of a strong Comptonized soft excess during a high-accretion state and its significantly weaker or negligible presence during a low-accretion state over a time span of $\sim$3~years \citep{jana2021, tripathi2022}. Similar trends are also exhibited by the flaring CLAGN LCRS B040659.9$-$385922, detected with eROSITA, \citep{krishnan2024} where the soft excess strengthened during the rise and faded during the decay of a $\sim$1.4~year flare. Overall, the soft excess in these transient individual CLAGN sources, including Mrk 1018, exhibits a direct correlation of the soft excess' intensity with an extremely variable accretion rate ($\lambda_{\rm Edd}$).

An alternate, although speculative, explanation of the disappearance of the warm corona and soft excess could be the destruction of the warm corona, driven by the strong optical outburst observed in 2020 \citep{brogan2023,lu2025}. However, we would need additional X-ray spectral studies of Mrk~1018 in the near future to adequately test if the soft excess has returned or remains absent, and thus to illuminate the connections between the presence or disappearance of the warm corona, long-term accretion rate changes, and the potential impact of the 2020 outburst. We note a partially analogous situation to that in the CLAGN 1ES~1927+654 \citep{trakhtenbrot2019}, where the hot corona was temporarily destroyed by an extrinsic event, possibly due to the impact of a tidally-disrupted star \citep{ricci2020}.

\subsection{The inner hot flow in Mrk 1018}\label{disc:hot-corona}
The hot Comptonized X-ray emission in Seyferts and Black Hole X-ray Binaries (BHXRBs) is proposed to originate from a hot-inner accretion flow, either in a compact X-ray corona \citep{haardt1991, haardt1993} or an advection dominated flow \citep{ptak1998} through upscattering of UV or optical seed photons. The seeds originate primarily from the optically thick accretion disk \citep[e.g.][]{done18}. 

\begin{figure}
    \includegraphics[scale=0.57]{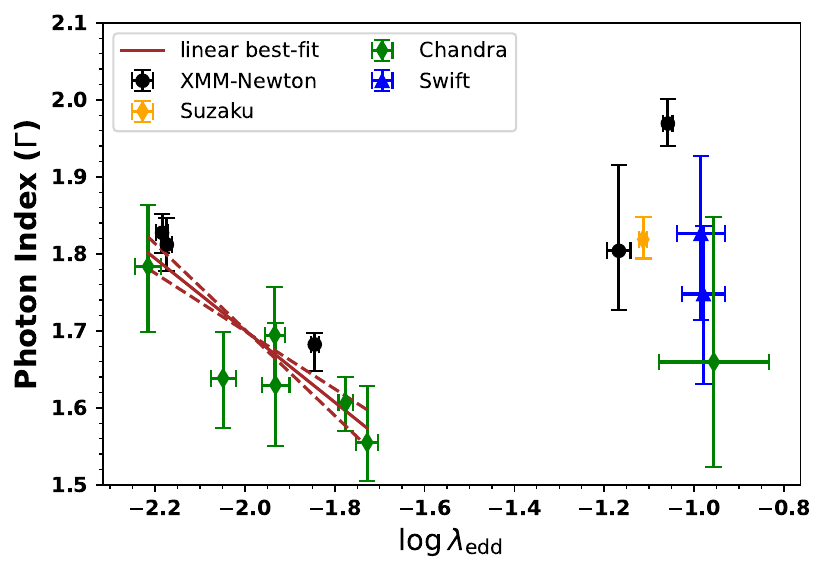}
    \caption{Evolution of photon index ($\Gamma$) of the intrinsic hot corona power law with respect to Eddington ratio ($\lambda_{\rm edd} = L_{\rm bol}/L_{\rm edd}$). The different colors represent different missions: green--{\it Chandra}, red--{\it Chandra} observation from 2010 evaluated from readout streak, blue--\textit{XMM-Newton}, orange-- \textit{Suzaku}, brown solid line: linear regression best-fit in the `softer when fainter' state, brown dashed line: the uncertainty range of the linear fit. The trend soft-when-fainter trend is evident for $\log \lambda_{\rm Edd}<-1.7$.}\label{fig:flux_vs_gamma}
\end{figure}

A softer-when-brighter trend for accretion rates above $\log(\lambda_{\rm Edd}) \sim -2$ has been noted in samples of Seyferts as well as in individual BHXRBs \citep[e.g.][]{lusso2010}. Meanwhile, at lower accretion rates, a softer-when-fainter behavior has been recorded for samples of low-luminosity AGN \citep{gu2009, younes2011, she2018} and in individual BHXRBs \citep[e.g. GRO J1655$-$40,][]{sobolewska2011b} in their low/hard state \citep{wu2008, yang2015}. We plot the hot corona photon index ($\Gamma$) vs Eddington ratio ($\lambda_{\rm Edd}$) for Mrk 1018 in Fig.~\ref{fig:flux_vs_gamma}. While in the bright phase the scatter in the data points does not reveal any trend. However, Mrk 1018 exhibits a `softer-when-fainter' state trend or an anti-correlation in the $\Gamma$-$\lambda_{\rm Edd}$ relation ($\rho_{\rm corr} = -0.89$, $p$=$1.4 \times 10^{-3}$) in the faint state after 2016. A linear regression returns the best-fitting relation

\begin{equation}\label{eq:gamma-lamdaedd}
    \Gamma = (-0.45 \pm 0.09) \log \lambda_{\rm Edd} + (0.79 \pm 0.16).
\end{equation}

Here, we have used only the high-count data and a physically-motivated model (\texttt{M3}), thus reducing
 the scatter in $\Gamma$ and the statistical uncertainty.
\cite{lyu2021} have reported this phenomenon in terms of $\Gamma$ and the X-ray flux (2--10~keV) for the low count {\it Swift}-XRT datasets, using only a simple power law fitting. 
Meanwhile, the UV-X-ray spectral index $\alpha_{\rm OX}$  in Mrk 1018 decreased initially but increased after 2018 (Table \ref{tab:agnsed}), a behavior similar to that of $\Gamma$. The `softer-when-fainter' trend in optical-X-ray spectral index ($\alpha_{\rm OX}$) is seen in samples of low-luminosity and changing look AGN \citep[e.g.][]{li2017, ruan2019}.
The two convergent trends in $\Gamma$ and $\alpha_{\rm OX}$ as a function of $\lambda_{\rm Edd}$ observed in individual BHXRBs, samples of CLAGNs and low-luminosity AGNs generally support the notion that a geometrically-thin disk dominates the spectrum at Eddington ratio values $\log \lambda_{\rm Edd} \gtrsim$ $-$2, whereas an ADAF dominates the accretion structure at values of $\log \lambda_{\rm Edd}\lesssim$  $-$2, respectively \citep[e.g.][]{sobolewska2011b,ruan2019}.
It would seem that the innermost accretion
structure in Mrk 1018 also underwent a transition as noted by \cite{lyu2021} as $\log \lambda_{\rm Edd}$ dropped below roughly $-$1.7. \citep{li2017}. {The observed timescale for $\Gamma$ to transition from softer-when-brighter to softer-when-fainter, hereafter $t_{\rm \Gamma}$, is thus indicative of (a) the timescale of disintegration or disappearance \citep[e.g.][]{hagen2024} of the optically-thick inner flow and/or (b) the timescale of over which the alternative seed photon source e.g. an ADAF producing synchrotron seed photons \citep[][and reference therein]{gu2009} dominates.}
As shown in Fig. \ref{fig:lc_all}c, $t_{\rm \Gamma}$ spans 4--10 years. The large uncertainty on $t_{\rm \Gamma}$ is due to the large uncertainty on $\Gamma$ in the 2010 \textit{Chandra} observation (see Fig.~\ref{fig:lc_all}c). If the value of $\Gamma$ had already fallen by 2010, this drop would be remarkable, since it would have occurred before the observed optical CLAGN transition, possibly indicating a precursory change in the accretion flow.

\subsection{Multiwavelength connection}\label{sec:multiwavelength}
\begin{figure*}
\centering
	\includegraphics[scale=0.5]{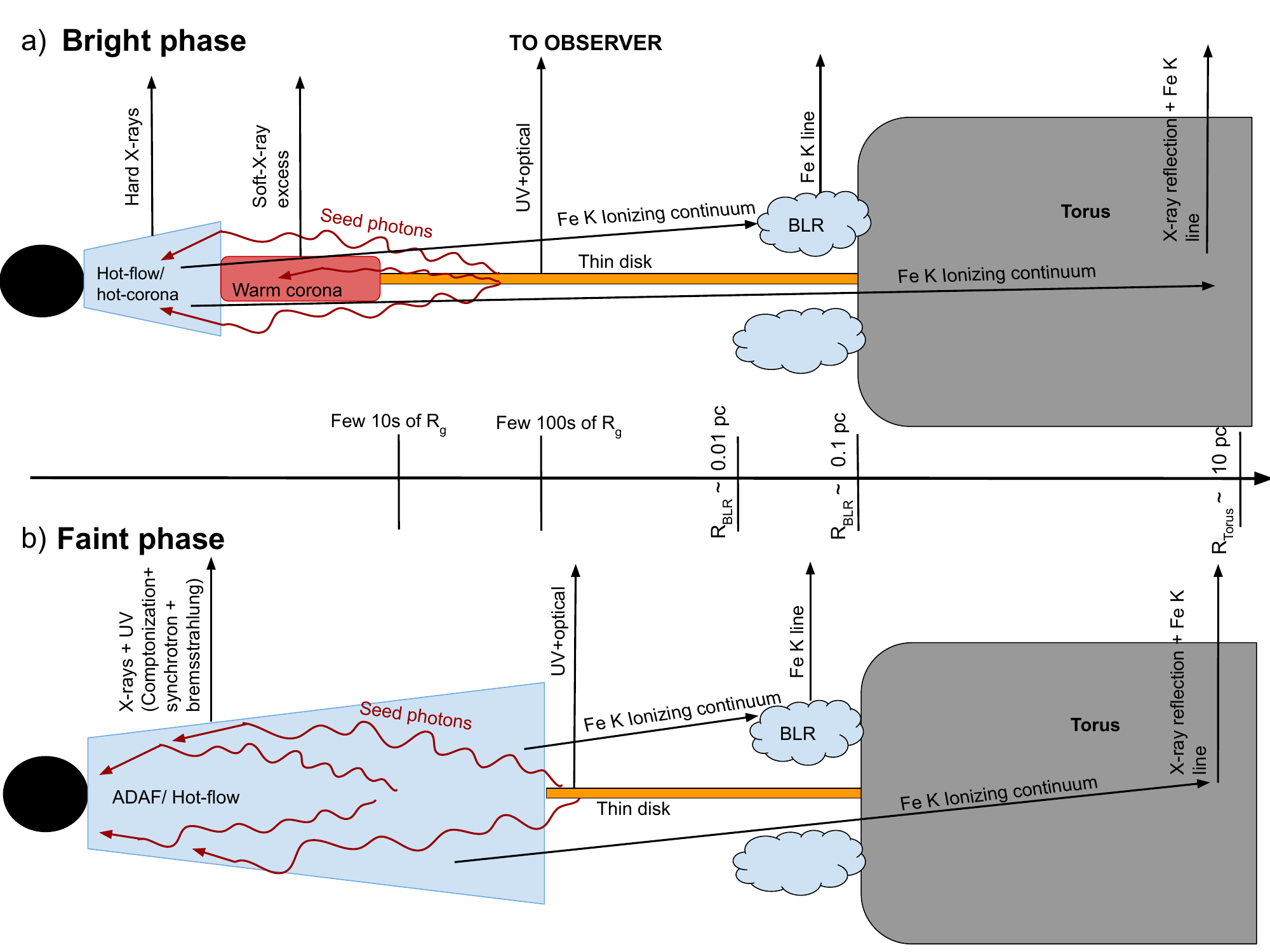}
\caption{Illustration of one possible geometrical change in AGN accretion structures due to the major CLAGN transition occurring after 2013. Length scales as illustrated here are approximate only. (a) bright state configuration: warm corona present in the inner accretion  structure along with the hot-accretion flow/corona/less dominant ADAF (b) faint state configuration after $\sim$2021: warm corona disappears (Sect.~\ref{disc:warm-corona}) and inner accretion flow dominated by a hot ADAF (Sect.~\ref{disc:hot-corona}), emitting X-rays and UV seed. The thin disk also decreases its energy contribution and physical size. Finally, the BLR+torus system extends from sub-parsec scale to 10s of parsecs, which results in significant smoothing of the Fe K$\alpha$ line response (Sect.~\ref{sec:fe-line-origin} and Fig. \ref{fig:fe_line_response}).}
\label{fig:type-transition-diagram}
\end{figure*}

Over the bright-to-faint transition from 2010--2013, the UV luminosity flux drops by a factor of $\sim$24 whereas the X-ray drops by a factor is $\sim$8 (Table \ref{tab:component-variability}). Additionally, the UV integrated model luminosity in the 2--15 eV band contributes a significant fraction to the UV--X-ray SED (Table \ref{tab:agnsed}) where $L_{\rm UV}/(L_{\rm UV} + L_{\rm X-ray})$ $\sim$0.6 to 0.7 in the bright phase and $\sim 0.5$ in the faint state, with $L_{\rm UV} > L_{\rm X-ray}$ for most broadband spectrum, similar to most Seyferts accreting at $\log \lambda_{\rm Edd}$ $\sim$ $-$2. Thus, from an energy argument, the larger UV variability-factor and its larger contribution to the total luminosity indicate that Comptonization of the UV seed-photons \citep{haardt1991, haardt1994} in a hot-corona/accretion flow is more likely to drive long term X-ray variability; as opposed to X-ray driving the UV via thermal reprocessing. This picture is also reinforced by other studies \citep[e.g.][]{uttley2003, hagen2023} explaining the inadequacy of the X-ray reprocessing in driving large amplitude-long term optical/UV variability in normal non-CL AGN. Additionally, UV emission driving X-ray emission is consistent with the notion that the physical process triggering the CLAGN transition initially affects a region in the accretion flow that primarily emits UV.

Furthermore, there is a non-trivial mismatch between the factors in the drops of UV and X-ray fluxes ($R_{\rm UV} \simeq 3R_{\rm HX}$; Table \ref{tab:component-variability}).
If the hot corona is static in morphology and physical properties across the bright and faint states, then the number of UV seed photons intercepted by the hot corona for upscattering to X-rays should decrease by the same factor as the UV flux (Appendix \ref{sec:seed-vs_xray}). In other words, the observed X-rays should also drop by that same factor. 
However, the X-rays drop by a much lower factor, ($R_{\rm HX}$) indicating that the drop in the number of seed photons is partially compensated by an increased level of Comptonization, and/or by other radiative processes in the faint state.
Such a situation can be achieved if the inner hot Comptonizing accretion flow becomes relatively more energetically dominant by (a) increasing its covering factor by a factor of 3, thus intercepting a higher fraction of seed photons in the faint state (3$\times$) than in the bright state, albeit with a lower total X-ray luminosity, (b) increase in the energy imparted per electron and seed-photon scattering due to a hotter corona/ADAF in the faint state and/or (c) increasing the X-ray emission due to other radiative processes which may occur in radiatively-inefficient flows, such as synchrotron, cyclotron, and/or bremsstrahlung emission \citep{narayan1995}.

All these observations — the severely reduced UV luminosity in the faint state, the
value of $R_{\rm HX}$ being less than that of $R_{\rm UV}$, the disappearance of the soft X-ray excess, the inversion of the $\Gamma-\lambda_{\rm Edd}$ relation as noted in this work and in \citet{lyu2021} — support the scenario in which a switch occured between two accretion configurations (following \citealt{esin1997} for BH XRBs; see also \citealt{lyu2021} for Mrk~1018). Specifically, in the bright state, a geometrically-thin, optically-thick, UV/optical-emitting disk is present (Fig.~\ref{fig:type-transition-diagram}a), supplying seed photons to both a hot hard X-ray-emitting flow and a warm soft excess-emitting corona for X-ray production. In the faint state, the geometrically-thin disk retreats to larger radii. Meanwhile, the inner hot flow/ADAF increases its presence spatially and energetically (Fig.~\ref{fig:type-transition-diagram}b), yielding a higher covering fraction to intercept more seed photons for Comptonization, and/or producing more emission via other processes — synchrotron, bremsstrahlung, etc. If this scenario holds, the UV band should also increase more significantly than the X-ray band if Mrk~1018 switches back to its bright state configuration in the future.

\subsection{Variability timescales and driving mechanism} \label{sec:accretion-timescales}
Variability in an accretion disk can be characterized by multiple timescales \citep{frank2002}. The dynamical timescale is given by $t_{\rm dyn} = \sqrt{R^3/GM_{\rm BH}}$. The thermal and viscous timescales are given by $t_{\rm th} = (1/\alpha) t_{\rm dyn}$ and $t_{\rm visc}=(H/R)^{-2}t_{\rm th}$, respectively. These timescales are dependent on the radial distance ($R$), the disk aspect ratio ($H/R$), and viscosity parameter \citep[$\alpha$,][]{shakura1973}. By estimating these timescales and comparing them with variability timescales measured in Mrk 1018, we can speculate on the possible origin(s) of the source's extreme variability.

The observed timescales associated with the long-term CLAGN transition of Mrk 1018 are
(a) $t_{\rm bright} \simeq$30~years, the observed time through which Mrk 1018 maintained its bright state (optical type 1), and 
(b) $t_{\rm sc} \simeq$1~year (Sect.~\ref{sec:flux_variability}), the transition timescale, which quantifies how fast the broadband flux drops after roughly MJD 55500--56200, eventually leading to the latest optical type transition.

Assuming a black hole mass of $M_{\rm BH}=10^{7.9}M_{\rm \odot}$ \citep{mcelroy2016}, radial distance values of $R$=50--100~$R_{\rm g}$ in the accretion flow, a viscosity value of $\alpha$=0.01, and disk aspect ratio of $H/R$=0.001 (thin disk), we obtain dynamical timescales of $t_{\rm dyn}$ $\simeq$ 1.6--4.6 days, thermal timescales of $t_{\rm th}$ $\simeq$ 0.44--1.24~years, and viscous timescales (for a geometrically-thin disk) of $t_{\rm visc} \simeq$ 180--500~years, respectively.
The thermal timescale is thus consistent with our measured value of $t_{\rm sc}$. However, neither $t_{\rm th}$ nor $t_{\rm visc}$ for a geometrically-thin disk are consistent with $t_{\rm bright}$. If we increase the disk aspect ratio ($H/R$) to 0.2, then the viscous timescales ($t_{\rm visc}$) get reduced to 10--30 years, consistent with the observed value of $t_{\rm bright}$. 
Our work is consistent with previous suggestions that geometrically-thick disks are compatible with observed CLAGN transition timescales, e.g., \citet{dexter2019} for magnetic pressure-supported disks.

The observed timescales of Mrk 1018's transition can also be qualitatively explained in the context of disk instability models.
The inner region of the accretion flow in AGN is typically dominated by radiation pressure \citep[e.g.][]{laor1989, noda2018}, resulting in an instability \citep{lightman1974, noda2018, sniegowska2020}. The instability heats the disk \citep{lightman1974}, increasing the disk aspect ratio ($H/R$, a geometrically thicker flow as proposed above) on the thermal timescale ($\sim$ $t_{\rm th}$), and the bright state, with an elevated accretion rate, begins. The bright state then sustains for a time period of $t_{\rm bright}$, explained by a reduced viscous timescale. The instability ends, and the disk reverts to its faint state on a thermal timescale ($t_{\rm th}$), thus explaining the observed value of $t_{\rm sc}$. To summarize, an intrinsic instability in the inner accretion flow can thus self-consistently explain the timescales connected to the drastic flux drop between 2010--2012 and consequently the optical type transformation detected in 2016. Our proposed mechanism of the CLAGN transition exploits the fact that the above mentioned disk instability process can induce a long term extreme variability flare intrisically, which is a parallel alternate to extrinsic driving mechanism like chaotic cold accretion \citep[CCA;][]{gaspari2013,gaspari2017,gaspari2020} as described by \citet{veronese2024}.

In addition to its long term variability behavior, Mrk 1018 exhibited two brief outbursts in 2016 \citep{krumpe2017} and 2020 \citep{brogan2023}, each lasting $\lesssim$1~year. The disk radiation-pressure instability scenario can also explain these relatively shorter timescales, as ranges of various model parameters \citep[e.g. Fig. 4 of][]{sniegowska2020} can yield extreme variability flares of various timescales. However, we argue that the dominant changes in the structure and energetics of the inner accretion flow (Sect.~\ref{sec:multiwavelength}) are primarily driven by the 30 year-long bright state, as it has generated at least $25$ times more energy than the two year-long outbursts (Appendix \ref{sec:energetics}).

\subsection{The Fe K$\alpha$ emission line}
\subsubsection{Origin}\label{sec:fe-line-origin}
The Fe K$\alpha$ lines in AGN can either originate in the inner accretion flow, strongly broadened under the influence of the black hole’s gravity \citep[e.g.][]{fabian1989, fabian2000} and/or in the distant matter such as the BLR \citep[e.g.][]{bianchi2008} or the parsec-scale torus \citep[e.g.][]{ricci2014a}, consequently yielding a narrow core.

We estimated the constraints on the Fe K$\alpha$ line width (Tab.~\ref{tab:fluxes-photon-indices}) and thus constraining the inner radial bound of its origin. Our Gaussian modeling of the Fe K$\alpha$ (Sect.~\ref{sec:iron-line-analysis} and \ref{sec:chandra-analysis}) shows that the emission line has an average width of $\sigma_{\rm FeK\alpha}$ $\simeq 125^{+150}_{-80}$~eV over all phases. Assuming Keplerian motion ($R = GM_{\rm BH}/v^2$), the FeK$\alpha$ line width yields an inner radius value in the range 0.002~pc to 0.08~pc.

In samples of unobscured and mildly-obscured Seyferts \citep[e.g.][]{yaqoob2004, shu2010}, the narrow Fe K$\alpha$ line emission can be localized to the outer accretion disk or BLR, using {\it Chandra} HETG. In  NGC 4151, Fe K$\alpha$ profile decomposition using high-resolution spectroscopy \citep{xrism2024} yields that a significant fraction of the Fe K$\alpha$ line flux originates from gas commensurate with the BLR. Furthermore, reverberation mapping studies for NGC 4151 and NGC 3516 \citep{zoghbi2019, noda2023} also infer Fe K$\alpha$ line-emitting gas to be commensurate with the BLR. The torus also contributes a narrow line core and a Compton shoulder \citep{yaqoob2012, xrism2024} to the line profile. 
From the continuum estimates obtained from our broadband SED (Sec. \ref{sec:broad-band}) and the radius-luminosity expressions of \cite{bentz2013} and \cite{kaspi2007}, we obtain an estimate for the BLR size of $\sim$0.03~pc and $\sim$0.01~pc in the bright and faint state, respectively. This estimate is also consistent with the average BLR distance estimate of 0.02--0.05~pc from \citet{mcelroy2016} and \citet{lu2025}. The dust-sublimation radius for Mrk 1018 was estimated to be $\sim 0.1$~pc (faint state) by \citep{brogan2023, lu2025}. Thus, the range of the inner radius obtained from the Fe K$\alpha$ line width overlaps with the range demarcated by the BLR distance estimate and the dust-sublimation radius of Mrk 1018.

\subsubsection{\it The extent of the Fe K$\alpha$ line emitting region}
\begin{figure}
    \centering
    \includegraphics[scale=0.60]{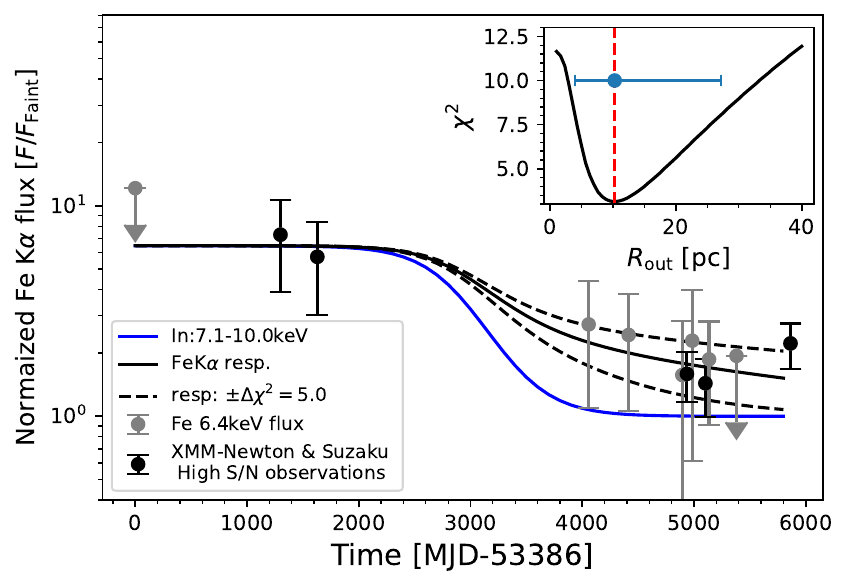}
    \caption{Response of an extended parsec scale structure with inner radius of $R_{\rm in}=0.01$~pc line response and outer radius $R_{\rm out} = 10$~pc. The dashed lines correspond to the upper and lower errors of $R_{\rm out}$. Inset: reduced-$\chi^2$ from the Fe K$\alpha$ light-curve fit best estimate of external radius. The blue marker and red dashed line represent the best-fitting value and the errorbar corresponds to $\Delta \chi^2=5$.}
    \label{fig:fe_line_response}
\end{figure}

The most likely explanation for the observed mismatch between the flux-drop factors of the emission line, $R_{\rm Fe}$~$\sim$3.4, and its driving continuum, $R_{\rm ion, Fe}=$6.8, is that the Fe K$\alpha$ line has not fully responded yet to reach its expected faint-state flux, which should be $\rm F_{faint}( Fe~K\alpha) = F_{bright}(Fe~K\alpha)/R_{ion,Fe}$. We model and interpret these drops in flux using a phenomenological reverberation model. In this model, a bi-conically cutout reprocessor \citep[a commonly used geometry in torus modeling, e.g.][]{ikeda2009, balokovic2018} is illuminated by an ionizing continuum originating from a central X-ray point source. Our numerical scheme adds up the response from the different volume elements accounting for time-delays. Assuming a static inner radius of $R_{\rm in}$=0.01~pc (consistent with a BLR) and opening angle $\theta_{\rm o}=45^{\circ}$, we estimate the radial extent (outer radius $R_{\rm out}$) of the reprocessor that is required to smear and smooth the observed Fe line response of an ionizing continuum ($F_{\rm ion}(t)$ fit by the MRS function; Sect.~\ref{sec:flux_variability}). We fit the Fe K$\alpha$ observed light-curve re-normalized with its plausible type 1.9 flux ($\rm F_{\rm faint}( Fe~K\alpha)$) with our model predicted light-curve. This returns an optimal value of $R_{\rm out}=10^{+17}_{-6}$~pc (Fig. \ref{fig:fe_line_response}). This estimate is consistent with the estimate that the infrared torus extends from sub-parsec distances to distances up to $\lesssim$100~pc \citep{honig2013,honig2017,honig2019}. Thus spatial extent of the Fe K$\alpha$ emitting torus indicates, that the iron emission line would require about additional $\sim9$~years from now to fully respond to the flux drop that occurred between 2010--2012 and settle to the faint state value. Detecting this drop will require high signal/noise monitoring of the source over the next decade.

\begin{figure}
    \centering
    \includegraphics[scale=0.58]{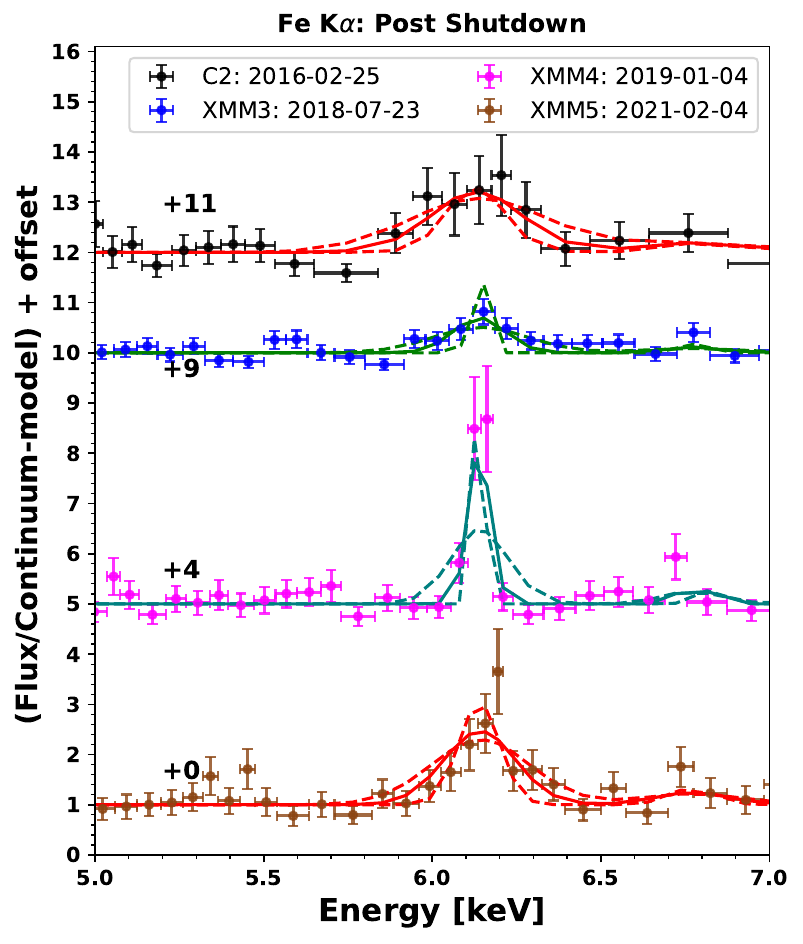}
    \caption{Spectral fits to the Fe K$\alpha$ line  from C2 (2016),  XMM3 (2018), XMM4 (2019), and XMM5 (2021) (datasets are grouped for clarity). The dashed lines indicate the profile corresponding to the upper and lower errorbars of $\sigma_{\rm FeK\alpha}$. The spectra here are normalized using the underlying continuum model. The relative strength/intensity of the Fe K$\alpha$ line with respect to the continuum is significantly highest for C2 and XMM5 and lowest for XMM3.}
    \label{fig:fe-line-ratio}
\end{figure}

This estimate of the outer radial extent of the iron line emitting region comes with a few caveats: (a) Our constraints on the inner radius ($R_{\rm in}$) are limited by CCD energy resolution, whereas high-resolution measurements are offered only by gratings and calorimeters. (b) There can potentially be some Compton scattering in relatively cold material of BLR/torus that might result in Compton-shoulders \citep[e.g.][]{yaqoob2011, buchner2019}, consequently affecting line width when modeling it with a single Gaussian. (c) The response of the Fe K$\alpha$ echo from a distant structure is based on a single scattering of the incident beam and does not take into account detailed radiative transfer effects. (d) The Fe K$\alpha$ flux from XMM5 which is included for the above analysis is taken immediately after the short 2020 outburst and might have minimal influence on the constraints obtained from long term trends.

Despite the limited dataset and the caveats mentioned above, we determined a reasonable estimate of the maximum path length covered ($\Delta r \sim$10~pc) by an ionizing continuum photon in the line emitting medium, indicating that a physically extended region (outer BLR to the parsec scale torus) likely acts to smear out the emission line response to the variable continuum. Additionally, our modeling strengthens our hypothesis that as of 2021, the Fe K$\alpha$ line has not undergone a full decay, and is still responding to the abrupt CLAGN `shutdown' event after 2013.

\subsubsection{Equivalent width and the line profile}
\begin{figure}
    \centering
    \includegraphics[scale=0.65]{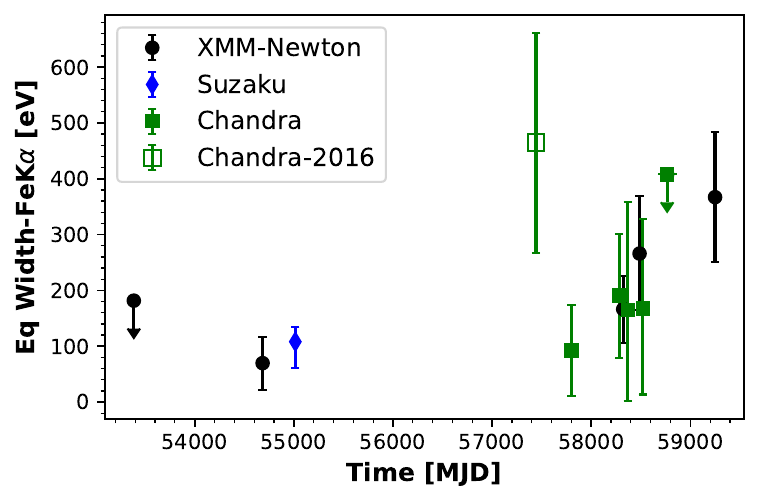}
    \caption{Time evolution of Fe K$\alpha$ equivalent width, which exhibits a mild increase with time. Black circular markers: {\it XMM-Newton}; blue diamond marker: {\it Suzaku}; unfilled green square marker: {\it Chandra} from 2016 (C2); green square markers: all other {\it Chandra} spectra.}
    \label{fig:eq-width}
\end{figure}
The Fe K$\alpha$ line in Mrk 1018 exhibits flux variations (Fig. \ref{fig:fe-line-ratio}), which consequently affects its equivalent width (EW).
A systematic increase in EW (excepting the 2016 observation) is observed as the flux drops. The trend exhibits a Pearson's correlation coefficient of $\rho_{\rm corr}=-0.62$ ($p=0.03$), indicating a mild anti-correlation. Such an anti-correlation is observed across multiple samples of type 1 and type 2 AGN \cite[e.g.][]{iwasawa1993, shu2012, ricci2014b, boorman2018}. The drivers of evolution in EW can be associated with two distinct physical processes: (a) variation of the underlying continuum \citep{jiang2006} and/or (b) variation of the covering fraction of distant narrow line-emitting material \citep{page2004} due to increasing radiation from the central engine \citep{ricci2017}. In this case, the Fe K$\alpha$ line originates in the sub-parsec BLR and/or the torus. Thus, a more relevant mechanism is the BLR covering fraction change induced by variations in $L_{\rm bol}$ on the thermal timescale ($t_{\rm th}$).
For our case, at $R=0.01$~pc (Sect.~\ref{sec:fe-line-origin}), $t_{\rm th}$ is as high as 160~years. Thus, our observed mild anti-correlation of Fe K$\alpha$ to the X-ray continuum does not result from the covering fraction change in the BLR/torus. The most plausible cause of the generic trend of EW variation is the slower drop of the Fe K$\alpha$ flux relative to the continuum X-ray flux.  

Considering observation C2, the Fe K$\alpha$ EW width exhibits a significant increase from pre-shutdown levels, $\sim$100~eV, to 464$\pm$200~eV in 2016 (Fig. \ref{fig:eq-width}), a distinct deviation from the long term increasing trend. We could not assign a particular cause to the sudden increase in EW in the C2 spectrum, except stochastic variability of the underlying continuum.

\section{Summary and conclusions} \label{sec:conclusions}

Mrk 1018 is part of the rare class of CLAGN that have exhibited multiple extreme variations in luminosity and optical spectral type transition. In the 1980s, Mrk 1018 transitioned from type 1.9/faint to a type 1.0/bright state \citep{cohen1986}. It remained in this bright state for roughly 30 years, after which it reverted to a type 1.9/faint state in the early 2010s \citep{mcelroy2016, husemann2016}.

Here, we analyzed the long-term multi-wavelength properties of the source (across 2003--2021), with a focus on the behavior during the transition from the bright state to the faint state, which started roughly between 2010 and 2013. In particular, we modeled all available high signal/noise X-ray spectra with a physically-motivated, multi-component model, and we quantified the broadband SED variability behavior.  We established the following characteristic changes in the inner accretion flow due to the CLAGN transition:

\begin{itemize}
\item The soft excess became fainter through 2019, maintaining temperatures between 0.1 and 0.2 keV and optical depths between 10--30.  However, the soft excess was not detected in the high signal/noise 2021 faint-state spectrum. It is possible that during the faint state, the warm corona structurally disintegrated, or became energetically negligible, by 2021, likely driven by the overall decrease in accretion rate.  Alternatively, the deficit of soft excess emission might be simply time-localized, driven by some disruption resulting from the 2020 optical/UV outburst.

\item The hot corona photon index ($\Gamma$), when plotted against $\lambda_{\rm Edd}$, follows a ``softer-when-fainter" trend in the faint state ($\log \lambda_{\rm Edd} \lesssim 10^{-1.7}$), as determined by the high-quality spectra (Eq.~\ref{eq:gamma-lamdaedd}). Our results are consistent with the $\Gamma$--$L_{\rm X-ray}$ trends obtained from the fitting of low signal/noise data by \cite{lyu2021} and that observed in some BHXRBs and samples of low-luminosity AGNs.

\item The broadband spectrum, from UV to X-rays, hardened during transition from the bright to faint state, as different emission components responded differently; the UV emission drops by a factor that is three times greater than the corresponding drop in hard X-ray emission.

\item The substantial drop in UV emission during the transition and the inverted $\Gamma$--$\lambda_{\rm Edd}$ relation observed during the low state are consistent with the CLAGN transition being driven by structural changes in the inner accretion flow. Specifically, the inner geometrically-thin disk transforms into a hot ADAF.  The observation that the hard X-ray flux from the hot corona drops by a factor lower than the UV thermal disk emission could be explained by either an increased covering fraction of the ADAF, or increased contributions from other X-ray continuum emission processes associated with ADAFs, such as synchrotron or bremsstrahlung.

\item While the Fe K$\alpha$ line flux is driven by variability in the hard X-ray continuum, the Fe K$\alpha$ line and the ionizing continuum exhibit significantly different variability factors. The observed line width and simple reverberation modeling constrain the extent of its emitting region to be between $0.01$~pc and a few tens of parsecs.

\item Furthermore, we predict that the Fe K$\alpha$ line has not fully responded as the more distant regions of the line-emitting gas are yet to respond. Simple modeling predicts that it could still require an additional $\sim$9 years from now for the Fe K$\alpha$ to settle down to a new equilibrium value.

\end{itemize}

The complex multiwavelength and multi-component behavior
in Mkn 1018 warrants continued long-term monitoring across the EM spectrum.
If the extreme variability observed in Mrk~1018 results from a limit-cycle behavior, then multiple successive short and long-term outbursts (with a few outbursts already observed) are a possibility. 
Speculatively, such behavior may also be occurring within the class of objects exhibiting recurrent CLAGN events, and Mrk~1018 may be a part of this class \citep{wang2024}.  Studying such outbursts, both in samples and in individual objects, will allow the community to progress theoretical models of accretion instabilities.

 Furthermore, integral field or mm-wave observations (e.g., ALMA) can map out cold circumnuclear structures, testing if substantial reservoirs of cold gas remain to fuel future bright phases, as well as further testing the viability of cold chaotic accretion (CCA) model in triggering CLAGN transitions.

\begin{acknowledgements}
The authors thank the anomymous referee for the comments and suggestions.

TS acknowledges full and partial support from Polish Narodowym Centrum Nauki grants 2018/31/G/ST9/03224, 2016/23/B/ST9/03123, and from Deutsches Zentrum für Luft- und Raumfahrt (DLR) grant FKZ 50 OR 2004.
MK is supported by DLR grant FKZ 50 OR 2307.
AM acknowledges full or partial support from Polish Narodowym Centrum Nauki grants 2016/23/B/ST9/03123, 2018/31/G/ST9/03224, and 2019/35/B/ST9/03944.
MG acknowledges support from the ERC Consolidator Grant \textit{BlackHoleWeather} (101086804).

This research is based on observations obtained with XMM-Newton, an ESA science mission with instruments and contributions directly funded by ESA Member States and NASA.  

This research has made use of data obtained from the Chandra Data Archive provided by the Chandra X-ray Center (CXC).

This research has made use of data and/or software provided by the High Energy Astrophysics Science
Archive Research Center (HEASARC), which is a service of the Astrophysics Science Division at NASA/GSFC. 

This research has made use of the NASA/IPAC Extragalactic Database (NED), which is funded by the National Aeronautics and Space Administration and operated by the California Institute of Technology.

We acknowledge the use of public data from the Swift data archive.

This work made extensive use of the following \texttt{python} packages: \texttt{NumPy} \citep{numpy}, \texttt{Matplotlib} \citep{matplotlib}, \texttt{SciPy} \citep{scipy}, \texttt{pandas} \citep{pandas1, pandas2}, and \texttt{Astropy}\citep{astropy2022}.

\end{acknowledgements}

\bibliographystyle{aa}
\bibliography{aa-mrk1018}

\begin{appendix}

\section{Mrk 1018 flux correction}\label{apdx:lc-corr}

\begin{figure}[H]
    \centering
    \gridline{\fig{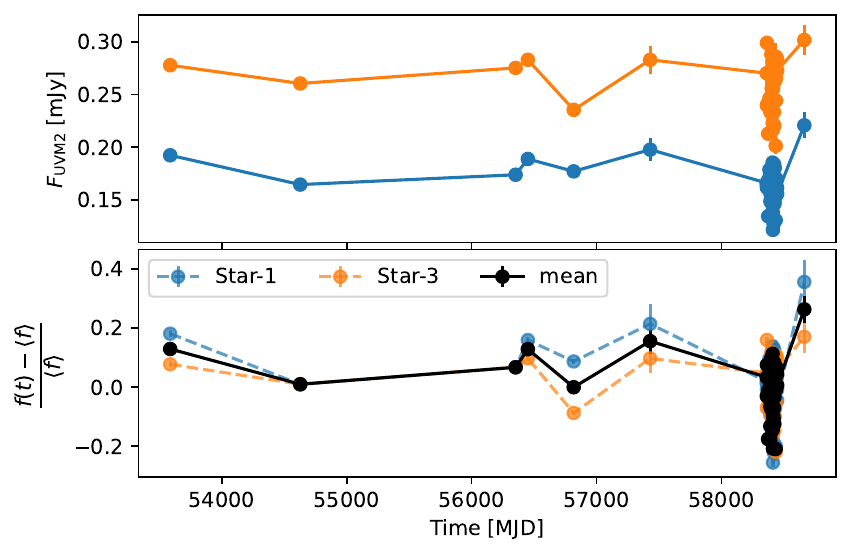}{0.42\textwidth}{}}
    \gridline{\fig{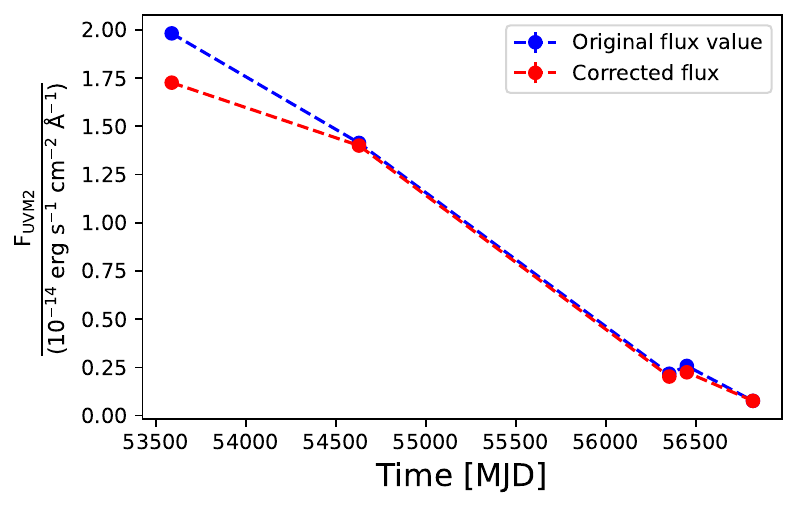}{0.42\textwidth}{}}
\caption{Top panel: stellar light-curve and time variable flux-correction factors. Bottom panel: light curve of Mrk 1018 in the bright and transition phase,
before and after flux corrections are applied following the method in Appendix \ref{apdx:lc-corr}.}\label{fig:lc-corr}
\end{figure}

We observed STAR-1 and STAR-2 \citep{brogan2023} using {\it Swift}-UVOT.
For each star, the average flux uncertainties was 0.002~mJy and 0.004~mJy, respectively.
The standard deviation of the flux measured at different epochs is $\sim$0.02~mJy for each star.
Thus, for both stars we find that the flux measurement uncertainties do not explain the observed flux variability.
The reasons for such variability cannot be particularly assigned.
However, we assume that the same variability trend is superposed on the AGN variability of in Mrk 1018, requiring us to correct the AGN source light curve, $F_{\rm AGN}(t)$.
We adopt a non-parametric approach to correct the Mrk 1018 {\it Swift}-UVOT fluxes for the epochs before MJD $\sim$57000.
The actual non-variable flux ($f_{\rm 0}$) of a given standard star is well represented by the mean of the variable flux $\langle f \rangle$, assuming that the variability in excess of the error bars is purely stochastic and not systematic.
Thus, the flux deviation factor for each star at a give epoch is
\begin{equation}
    C_{\rm STAR}(t) = \frac{f(t) - \langle f \rangle}{ \langle f \rangle}
\end{equation}

The average deviation factor for the two stars are thus $C_{\rm avg}(t) = [C_{\rm STAR-1}(t) + C_{\rm STAR-3}(t)]/2$ (Fig. \ref{fig:lc-corr}a).
The correction of flux of Mrk 1018 as derived from the deviation factor is thus given by $\Delta F_{\rm corr, AGN}(t) = C_{\rm avg}(t) F_{\rm AGN}(t)$.
The corrected flux of Mrk 1018 is thus given by 
\begin{equation}
    F_{\rm corr, AGN}(t) =  F_{\rm AGN}(t) [1 - C_{\rm avg}(t)].
\end{equation}
The maximum deviation in the corrected AGN flux with respect to the originally observed flux was found to be $\sim$0.3~mJy, affecting the first UVM2 data point. Overall, the corrected flux in Fig. \ref{fig:lc-corr} drops from (1.72$\pm$0.05)$\times 10^{-14}$~erg~cm$^{-2}$~s$^{-1}$~$\AA^{-1}$ during MJD 53587.0 to (0.07$\pm$0.01)$\times 10^{-14}$~erg~cm$^{-2}$~s$^{-1}$~$\AA^{-1}$ during MJD 56817.0.

\section{Ratio spectrum calculation}\label{sec:ratio-spec}
The average model spectrum in the bright/faint state is defined as
\begin{equation}
    F_{\rm state}(E) =  \frac{\sum_i F_{\rm state,i}(E)}{N_{\rm state}}
\end{equation}
where $F_{\rm state, i}(E)$ is the evaluated model spectrum at the best-fit parameters from the \texttt{M3} model fits (Sect.~\ref{sec:xmm-suzaku-analysis} and Table \ref{tab:xmm-suz-parameters}) for either bright or faint state. The error bar, $\sigma_{\rm state}(E)$, at each defined flux value here is the 90\% confidence level calculated by evaluating the model on a subset of 500 samples selected randomly from the parameter-posterior distribution.
The flux ratio at a given energy is defined by
\begin{equation}
    R(E) = F_{\rm bright}(E)/F_{\rm faint}(E)
\end{equation}
where the average in the `bright' state is taken over XMM1, XMM2, and SUZ, and the `faint' state is taken over XMM3, XMM4, and XMM5. The uncertainty in the ratio is given by
\begin{equation}
    \sigma_R(E) = R(E) \sqrt{ \left( \frac{\sigma_{\rm bright}(E)}{F_{\rm bright}(E)} \right)^2 +  \left( \frac{\sigma_{\rm faint}(E)}{F_{\rm faint}(E)} \right)^2}
\end{equation}

The UVM2 bright to faint state flux density ratio ($R_{UVM2}$) is calculated using the data point from the {\it XMM-Newton} observations. The energy band for UVM2 is $E_{\rm UVM2}=(5.52 \pm 1.52)\times 10^{-3}$~keV. 

The ratios $R(E)$ and $R_{\rm UVM2}$ are plotted in Fig. \ref{fig:spectra_change_between_types}.

\section{UV seed and X-ray variability}\label{sec:seed-vs_xray}
Here, we describe how the extreme variability in an accretion disk is transferred to a static lamp-post corona. This exercise helps us understand the relation between the flux drop factors in the UV band vs.\ that of the X-ray band.
The hot-corona X-ray emission is the up-scattered emission from the accretion disk. The variable disk emission at a given radius can be modeled by the emission from a thermal accretion disk \citep{shakura1973}, given by
\begin{equation}
    F_{\rm disk}(t,r) = \frac{3GM_{\rm BH} \dot{M}(t)}{16\pi r^3} \left( 1- \sqrt{r_{\rm isco}/r} \right)
\end{equation}
In the above expression, $\dot{M}(t) = \frac{\lambda_{\rm Edd}L_{\rm Edd}}{\eta c^2}
f_{\rm long}(t)$, where all parameters are constant and the variabilily is introduced by the modified sigmoid function (MRS), $f_{\rm long}(t)$ (Sect.~\ref{sec:flux_variability}). Here the efficiency factor $\eta$ is 0.057 as adopted by \cite{done18}, $\lambda_{\rm Edd}$ is 0.02, and $M_{\rm BH}=10^{7.9}M_{\odot}$. The parameters of the MRS functions are $R=24.5$, $t_{\rm 0}=2200$~days, and $t_{\rm sc}=365$~days (Sect.~\ref{sec:flux_variability}).

Here, we consider the dominant UV emitting inner disk region to be between $r_{\rm trunc}=30~R_{\rm g}$ and $r_{\rm o}=70~R_{\rm g}$. The variable accretion rate $\dot{M}(t)$ wavefront begins at an instability location $r_{\rm o}$ and travels inwards ($r<r_{\rm o}$) at a finite speed $v$ down to the truncation radius $r_{\rm trunc}$.
While the disturbance propagates through each annulus it also transmits seed photon `signals' to the corona, which has a finite radius of $a$ and is located at a height $h_{\rm X}=30~R_{\rm g}$ above the disk-plane of the axis of symmetry.
The seed photon signal received by the corona is subject to delays due to
(a) the finite speed of $v$ ($< c_{\rm s}$, the speed of sound) of wave propagation in the disk, and
(b) the delay from the light-travel time for the distance $d(r)=\sqrt{h_{\rm X}^2 + r^2}$. The total delayed time of the wavefront is thus given by
\begin{equation}
t_{{\rm delay},t,r} = t + \frac{r-r_{\rm o}}{v} - \frac{d(r)}{c}
\end{equation}
The radiation intercepted by a hot spherical lamp-post corona from a disk annulus at radius $r$ is given by:
\begin{equation}
    dL_{\rm cor, inc}(t,r) = 2\pi r F_{\rm disk}(r, t_{{\rm delay},t,r}) \frac{\pi a^2 h}{[h^2_{\rm X} + r^2]^{3/2}}  dr
\end{equation}
The above expression is integrated in the range $r_{\rm trunc} \leq r \leq r_{\rm o}$. The total variable emission incident on the corona is thus given by $L_{\rm inc,cor}(t) = \int_{r_{\rm trunc}}^{r_{o}} dL_{\rm cor, inc}(t,r)$. 
We simulate the light curves signifying the emission intercepted by the corona for two different values of corona size, $a=2~R_{\rm g}$ and $a=5~R_{\rm g}$.

\begin{figure}
    \centering
    \includegraphics[scale=0.7]{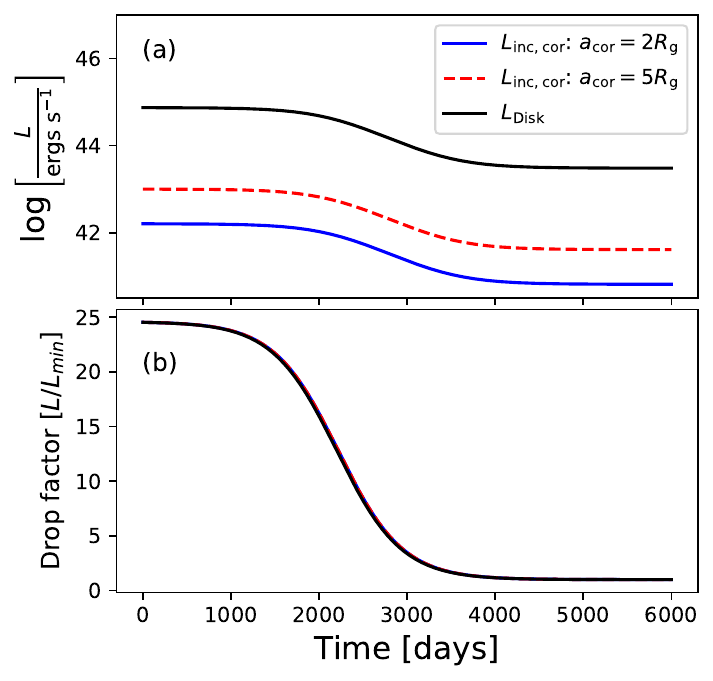}
\caption{(a) Top panel: incident seed luminosity ($\log L_{\rm inc,cor}$) on a corona of radius $2R_{\rm g }$ (blue), radius $5R_{\rm g}$(red) and the accretion disk luminosity ($L_{\rm disk}$) (b) Bottom panel: Same light-curves but normalized using the minimum luminosity value. The drop factors are equal.}\label{fig:lc-cor-disk}
\end{figure}

The variability light curves  of the intercepted seed photon ($L_{\rm inc, cor}$) and the total disk emission ($L_{\rm Disk}$) between $r_{\rm trunc}$ and $r_{\rm o}$ (Fig. \ref{fig:lc-cor-disk}) show that for two given corona sizes which are constant in time the net amount of intercepted seed photons varies significantly, with no difference in the variability factor for any light curves.
There is also negligible difference in the variability profiles since the net delays are not enough to smooth the light curves on timescales of $t_{sc} \simeq 1$~year.
This indicates that we require a increased covering fraction of the hot-X-ray emitter  with respect to the UV emitter i.e. variable $h_{\rm X}$, corona size $a$, in the faint state to obtain different values of long term flux drop in the UV and the X-ray waveband.

As a caveat, we note that the disk-corona model assumed here has a simple geometry. However, the phenomena reported above would be largely geometry independent. This is because the results depend on the physical extent of the corona with respect to the accretion disk, and our model effectively captures this aspect. Furthermore, we do not assume an inner ADAF despite the fact that the inner flow structure changes across accretion rates.

\section{Energy budget during different phases of Mrk 1018's evolution}\label{sec:energetics}
We have fit a modified reverse sigmoid function (Sect. \ref{sec:flux_variability}) to the X-ray flux data post 2005. The best-fit light-curve has captured the long term flux variability trend including the major flux drop post 2013. 
We can then apply the X-ray bolometric corrections from \citet{duras2020} to the long-term X-ray light curve $L_{\rm X-ray}(t)$) to obtain the long-term bolometric luminosity light-curve, $L_{\rm Bol}(t)$.
In the bright state, the excess luminosity generated compared to the faint state $\Delta L_{\rm Bol}(t)= L_{\rm Bol}(t) - L_{\rm Bol, faint}$.
Thus, the excess energy emitted in the bright state compared to the faint state is $\Delta E_{\rm bright} = \int \Delta L_{\rm Bol}(t) dt$. Integrating from 2005 to 2015, we obtain $\Delta E_{\rm bright,2005-2015} = 2\times10^{53}$~ergs. Assuming that the bright state had started in 1986 and ended in 2015 \citep{lu2025}, i.e. continued for 30 years, and that the source has maintained the same average bright state flux in this period, the total power generated in the 30 year bright state phase is $\Delta E_{\rm bright-phase} \simeq 3\Delta E_{\rm bright,2005-2015} = 6 \times 10^{53}$~ergs. We can convert this estimate to the equivalent amount of mass accreted using  $\Delta M_{\rm bright-phase}=\Delta E_{\rm bright-phase}/\eta c^2$, where accretion efficiency $\eta$ is taken to be 0.057 \citep{done18}. We obtain $\Delta M_{\rm bright-phase}$ $\simeq$6$M_{\rm \odot}$.

We can also estimate roughly the energy emitted in the 2020 outburst \citep{brogan2023, lu2025}. The outburst extends for a comparatively short time, 1~year. If we assume that it has same peak luminosity as that of the pre-2010 bright state, then it should have a total energy output of $E_{\rm outburst, 2020} \sim 2.5 \times10^{52}$~ergs ($\Delta M_{\rm outburst}=0.24M_{\rm \odot}$). This estimate is approximately 25 times less than $\Delta E_{\rm bright-phase}$, the energy emitted in the 30-year long `flare' in the bright phase.

\end{appendix}
\end{document}